\documentclass[11pt,a4paper]{article}
\usepackage{multicol}
\usepackage{comment}
\usepackage{booktabs}
\usepackage{tabularx}
\usepackage{multirow}
\usepackage{colortbl}
\usepackage{siunitx}
\usepackage{axodraw4j}
\usepackage[dvipsnames]{xcolor}
\usepackage{jheppub}
\usepackage{caption, subcaption}  
\usepackage{color,latexsym, array,multirow,  verbatim, enumerate,cancel,graphicx}
\usepackage{slashed}
\usepackage{bm}

\def\beq {\begin{equation}}
\def\eeq {\end{equation}}
\def\bi {\begin{itemize}}
\def\ei {\end{itemize}}
\def\bea {\begin{eqnarray}}
\def\eea {\end{eqnarray}}

\numberwithin{equation}{section} 

\title{LLPNet: Graph Autoencoder for Triggering Light Long-Lived Particles at HL-LHC}

\author{Biplob Bhattacherjee$^1$, Partha Konar$^2$, Vishal Singh Ngairangbam$^2$,  Prabhat Solanki$^1$ }

\affiliation{\vspace*{0.1in}$^1$ Centre for High Energy Physics, Indian Institute of Science, Bengaluru 560012, India}
\affiliation{\vspace*{0.1in}$^2$ Theoretical Physics Division, Physical Research Laboratory, Shree Pannalal Patel Marg, Ahmedabad, 380009, Gujarat, India}

\emailAdd{biplob@iisc.ac.in}
\emailAdd{konar@prl.res.in}
\emailAdd{vishalng@prl.res.in}
\emailAdd{prabhats@iisc.ac.in}

\abstract{ In the search for exotic events involving displaced particles at HL-LHC, the triggering at the level-1 (L1) system will pose a significant challenge. This is particularly relevant in scenarios where low mass long-lived particles (LLPs) are coupled to a Standard Model (SM)-like 125 GeV Higgs boson and they decay into jets. The complexity arises from the low hadronic activity resulting from LLP decay, and the existing triggers' inability to efficiently select displaced events. This study introduces a novel machine learning approach to address this challenge, utilizing a lightweight autoencoder architecture designed for low latency requirements at L1. Focusing on light LLPs with decay lengths ranging from 1 to 100 cm, this approach employs ``Edge convolution" on L1 reconstructed tracks. The results show notable signal acceptance at the permissible background rate, primarily originating from minimum bias and QCD di-jet events. For LLPs of mass 10, 30, and 50 GeV at decay length of 5 cm, the signal efficiencies are 33\%, 70\%, and 80\%, respectively. At a 50 cm decay length, these efficiencies are 20\%, 39\%, and 45\% for the same respective masses.
}

\begin{document}
\maketitle
\flushbottom


\section{Introduction}
\label{sec:intro}
The Large Hadron Collider (LHC) is yet to discover any concrete signal of new physics, with most measurements of the recently discovered Higgs boson pointing towards its agreement with the Standard Model. With other observations pointing toward the need for beyond-(the)-Standard-Model (BSM) physics, it is important to be thorough while looking for possible deviations from known physics at the LHC. One such signature traditionally overlooked is displaced signature. Recently, there has been a growing interest in investigating these signatures from both a phenomenological and experimental perspective. On the phenomenological side, various studies have investigated these signatures \cite{Ilten:2015hya,Gago:2015vma,Banerjee:2017hmw,KumarBarman:2018hla,Bhattacherjee:2019fpt,CidVidal:2019urm,Banerjee:2019ktv,Jones-Perez:2019plk,Bhattacherjee:2020nno,Bhattacherjee:2021qaa,Adhikary:2022goh,Gershtein:2020mwi,Fuchs:2020cmm,Cheung:2020ndx,Liu:2020vur,Evans:2020aqs,Linthorne:2021oiz,Alimena:2021mdu,Bhattacherjee:2021rml,Sakurai:2021ipp,Du:2021cmt,Bhattacherjee:2023plj,Bhattacherjee:2023kxw}. Similarly, on the experimental side, LHC experiment has also focused on investigating these signatures  \cite{CMS:2014hka,CMS-PAS-EXO-16-022,CMS-PAS-EXO-16-036,CMS:2017kku,CMS:2018lab,CMS:2018bvr,CMS-PAS-FTR-18-002,CMS-PAS-EXO-19-013,CMS:2019qjk,CMS:2019zxa,CMS:2020atg,CMS:2020iwv,CMS:2021juv,CMS:2021kdm,CMS:2021yhb,ATLAS:2015xit,ATLAS:2015wsk,ATLAS:2015itk,ATLAS:2017tny,ATLAS:2018rjc,ATLAS:2018lob,ATLAS:2018niw,ATLAS:2018tup,ATLAS:2019qrr,ATLAS:2019kpx,ATLAS:2019fwx,ATLAS:2019tkk,ATLAS:2019jcm,ATL-PHYS-PUB-2019-002,ATLAS:2020xyo,ATLAS-CONF-2021-015,ATLAS-CONF-2021-032,ATLAS:2020wjh,ATLAS:2021jig,ATLAS:2023oti,LHCb:2016buh,LHCb:2016inz,LHCb:2016awg,LHCb:2017xxn,LHCb:2019vmc,LHCb:2020akw,Feng:2017uoz,Curtin:2018mvb,Alpigiani:2020iam,Gligorov:2017nwh,Aielli:2019ivi}. While much of the investigation into displaced signatures at the LHC has been conducted using detectors like CMS, ATLAS, and LHCb, these studies have primarily relied on level-1 (L1) triggers originally designed for prompt signatures. Nevertheless, there are exceptions, such as specific studies conducted by ATLAS, like the one mentioned in \cite{ATLAS:2019qrr}, which employed a dedicated trigger known as \textit{CalRatio} \cite{ATLAS:2013bsk} to identify events involving long-lived particles (LLPs) decaying into jets, either within the hadronic calorimeter (HCAL) or near the outer radius of the electromagnetic calorimeter (ECAL). In addition, some phenomenological studies have proposed specialized triggers for displaced jets, considering the unique characteristics of these jets, including their trackless-ness nature and their substantial time delay within the ECAL \cite{Bhattacherjee:2020nno,Bhattacherjee:2021qaa}. In this paper, we explore the possibility of utilizing lightweight anomaly detection methods for point clouds to design dedicated triggers to select LLP events at the L1 trigger system in future runs of the LHC.

Triggering events with displaced signatures will pose a major challenge at the L1 trigger system at the high luminosity LHC (HL-LHC) \cite{Apollinari:2017lan} where instantaneous luminosity will reach $\mathrm{5 \times 10^{34} cm^{-2}s^{-1}}$ with average number of pile-up (PU) interactions reaching up to 140. With $pp$ collision happening every 25 ns, event rate will be as high as 40 MHz which need to be brought down by a factor of $\mathcal{O}(10^3)$ at the L1 stage of the trigger system using triggers dedicated for the displaced physics searches. To deal with this challenge, the data acquisition process at L1 of the trigger system will undergo a significant overhaul through the upgrade of the inner and outer tracker and the implementation of Field Programmable Gate Arrays (FPGA) \cite{CERN-LHCC-2020-004}. This upgrade will allow tracking information to be available at L1. Consequently, the improved architecture for data acquisition and processing at L1 will facilitate the utilization of machine learning techniques, as well as higher-level object reconstruction, within the trigger system. This advancement will prove invaluable in capturing rare BSM events, particularly those involving displaced objects that would otherwise have been missed. Furthermore, by implementing extended tracking at L1, it will become possible to reconstruct displaced tracks up to a specific transverse impact parameter, enabling the identification of events exhibiting displaced signatures at L1 \cite{CERN-LHCC-2020-004}. 

Triggering events containing displaced particles using standard PUPPI \cite{Bertolini:2014bba,CERN-LHCC-2020-004} jet triggers might not be efficient as shown in \cite{Bhattacherjee:2021qaa}. Specifically, triggering events that involve displaced physics objects resulting from the decay of low-mass LLPs, such as LLPs originating from the exotic decay of 125 GeV Higgs boson \cite{Curtin:2013fra}, will pose a significant challenge at HL-LHC. In such scenarios, standard $H_T$ triggers will also be inefficient to capture these events effectively due to the limited hadronic activity in the calorimeters. Therefore, there is an urgent requirement to develop dedicated triggers specifically designed to efficiently select the events containing displaced objects while effectively controlling the background rate.

Deep learning~\cite{LeCun2015,Silver2016,devlin-etal-2019-bert} methods offer powerful ways to process the large and complicated data obtained at the LHC~\cite{PhysRevLett.65.1321,DENBY1988429,deOliveira:2015xxd,Kasieczka:2019dbj,Guest:2018yhq}, and anomaly detection techniques~\cite{Heimel:2018mkt,Farina:2018fyg,Blance:2019ibf,Roy:2019jae,Dillon:2022mkq,Anzalone:2023ugq} based on deep neural networks offer a powerful way of identifying displaced signatures at any point in the analysis chain at the LHC. With the hardware upgrades to FPGAs at the L1 stage and enhancements to the DAQ system at HL-LHC, the integration of unsupervised machine learning algorithms into the L1 trigger system now seems achievable \cite{Duarte:2018ite,Summers:2020xiy,Govorkova:2021utb,Neu:2023sfh}. In this paper, we introduce $\texttt{LLPNet}$--a lightweight graph autoencoder~\cite{Atkinson:2021nlt,Tsan:2021brw,Atkinson:2022uzb,Hao:2022zns} that utilizes edge convolutions ~\cite{wang2019dynamic} and can detect anomalous signatures of LLPs against the minimum bias and QCD di-jet backgrounds with the restricted tracker level information available at the L1 trigger.

Rest of the paper is structured as follows- In Section \ref{sec:sigback}, we outline the signal benchmark points and background. Section \ref{sec:network_arch} explains the network architecture of the autoencoder and the training procedure employed in this study. In Section \ref{sec:results}, we study the performance of the autoencoder for various signal benchmark points. Section \ref{sec:rate} presents the rate calculation for the various LLP scenarios and quantify results in terms of background rate and signal efficiency. Finally, we conclude and summarise in Section \ref{sec:summary}.  

\section{Signal and background}
\label{sec:sigback}

In this work, we explore the specific BSM scenario of LLPs coupling to the 125 GeV SM-like Higgs boson. The quest for LLPs within this model is substantially motivated given that the Higgs portal serves as one of the prominent renormalizable channels through which new gauge-singlet particles can interact with the SM \cite{Curtin:2013fra, Matsumoto:2018acr, Arcadi:2021mag, Bhattacherjee:2021rml}. Exotic signatures of the Higgs boson decaying into LLPs are prevalent in various BSM theories, the details of which are elaborated in \cite{Alimena:2019zri} and references therein. We consider decay of LLP into a pair of $b$-quarks with 100\% branching fraction. The whole process can be represented as-
\begin{equation*}
    \mathbf{pp \rightarrow h (125 GeV) \rightarrow XX, X \rightarrow jj} 
\end{equation*}

Where SM-like 125 GeV Higgs boson ($h$) is primarily produced through gluon-gluon fusion. It then decays to two LLPs ($X$), with each LLP subsequently decaying to two jets ($jj$). In this particular scenario, the mass of the LLPs is relatively small, with values below half the mass of the Higgs boson, making it challenging to trigger such events efficiently using existing triggering strategies. BSM searches looking for LLPs in such scenarios have to rely on the associated production modes of Higgs boson like vector-boson fusion (VBF) and Higgs-strahlung (VH) to trigger events at L1 using physics objects like jets and leptons produced in association with Higgs boson which comes at the cost of reduced cross-section. Our study focuses on LLPs with mass, $m_X=$ 10, 30 and 50 GeV with mean proper decay lengths ($c\tau$) spanning from 1 cm to 100 cm. Due to the small boosts of LLPs in this specific scenario, displaced jets originating from the decay of LLPs are soft and exhibit low hadronic activity within the detectors. Consequently, triggering on these LLPs becomes more challenging. For producing signal events and handling showering and hadronization, we use $\texttt{Pythia8}$ \cite{Sjostrand:2014zea}. We generate the signal samples using the $\texttt{CTEQ6L1}$ PDF (Parton Distribution Function) \cite{DanielStump_2003}, along with the $\texttt{CUETP8S1-CTEQ6L1}$ CMS tune \cite{CMS:2015wcf}.

The primary sources of background in the present study are ``minimum bias" (Minbias) events and QCD di-jet events. Minbias events refer to inelastic proton-proton ($pp$) collisions where momentum transfer is very low. Such events will be produced at the nominal HL-LHC luminosity with huge cross-section of approximately 80 millibarns (mb). These events contribute significantly to the overall background due to their huge cross-section. QCD di-jet events also contribute significantly to the background in the  current study. We divide the QCD background events in various ranges of parton level $p_T$ ($p_T^{gen}$) to generate sufficiently significant background events in different $p_T^{gen}$ bins along with minbias events. QCD events are generated in bins of $p_T^{gen} \in$ \{30, 50\}, \{50, 75\}, \{75, 100\}, \{100, 125\}, \{125, 150\}, \{150, 175\}, \{175, 200\} and $> 200$ GeV.  A large background dataset in the $p_T^{gen}$ bin, \{30, 50\} GeV, for training the autoencoder is produced using \texttt{MadGraph5\_aMC@NLO} \cite{Alwall:2014hca,Frederix:2018nkq} while showering is done using \texttt{Pythia8}. Simple detector simulation is carried out with \texttt{Delphes-3.5.0} \cite{deFavereau:2013fsa} for both the signal and background datasets. To mimic the high PU environment of the HL-LHC, we incorporate the effect of PU by merging 140 PU events\footnote{The average number of soft interactions is taken to be 140 following a Poisson distribution.} with each primary collision in \texttt{Delphes}. We made slight modifications to the \texttt{PileUpMerger} code in \texttt{Delphes} to properly account for the offset in the z-direction and timing values for the first stable particle which in current study comes from the decay of the LLPs. This adjustment ensures accurate positioning of the decay product of the LLPs and properly factor in the effects of PU in our analysis.

\section{Network architecture and training}
\label{sec:network_arch}
The convolution-like operation referred to as ``\texttt{EdgeConv}" (Edge convolution) was introduced for point clouds, as detailed in \cite{wang2019dynamic}. In this approach, the point cloud is interpreted as a graph, with individual points symbolizing nodes, and edges being formulated by linking each point to its closest neighboring points. The \texttt{EdgeConv} operation accepts a graph as input, where each node signifies a data point, and the edges describe the connections or relationships among those data points. Through the application of convolutional operations to the features of adjacent nodes in the graph, it generates updated features for every node. In essence, \texttt{EdgeConv} exploits the connectivity data between points within the graph structure of the point cloud, thereby allowing the operation to extract local information and conduct convolutions on the point cloud data.

Mathematically, the \texttt{EdgeConv} operation for the $l^{th}$ message passing operation can be represented as follows:

\[ 
\mathbf{h}_i^{(l+1)} = \Delta_{j \in \mathcal{N}(i)} \Theta^{(l)} \cdot (\mathbf{h}_i^{(l)}, \mathbf{h}_j^{(l)} - \mathbf{h}_i^{(l)},\mathbf{e}_{ij}^{(l)})
\]

where:
\begin{itemize}
    \item $\Delta$  is a permutation invariant aggregation operation, e.g., max, sum, or mean. 
    \item \(\mathbf{h}_i^{(l)}\) and \(\mathbf{e}_{ij}^{(l)}\)  are the input node and edge features (if present) at the $l$-th message passing operation, respectively.
    \item  \(\mathbf{h}_i^{(l+1)}\) represents the updated feature of node \(i\) at the \((l+1)\)-th layer.
    \item \(\mathcal{N}(i)\) denotes the set of neighboring nodes of node \(i\).
    \item \(\Theta^{(l)}\) denotes the learnable function at the \(l\)-th layer.
\end{itemize}

The \texttt{EdgeConv} operation captures the local relationships and information within the graph structure by considering the differences between the features of neighboring nodes. It enables the network to learn meaningful representations from the graph-structured data. 

In this study,  we focus on tracks reconstructed at L1 with a transverse momentum $p_T>2$ GeV and pseudorapidity $|\eta|<$ 2. Each of these tracks is treated as an individual node, while each event consisting of these tracks corresponds to the graph. Without any a priori given graph structure, we investigate two approaches to construct graphs--radius graphs and k-nearest neighbour ($k$-NN) graphs using the Euclidean distance in the spatial vertex in $x$, $y$, and $z$ coordinates. Radius graphs are constructed by connecting pairs of points within a specified radius (\( r_G \)) of each other--an edge is drawn between two nodes if their spatial separation is less than or equal to \( r_G \). The parameter \( r_G \) can be seen as a threshold that determines how close points must be to each other to be considered neighbors, while $k$-NN graphs are formed by connecting each point to its \( k \) nearest neighbors, regardless of the actual distances between those points. Here, \( k \) is a fixed integer representing the number of nearest neighbors to be considered. In radius graphs, we can control the granularity of the relationships by varying the radius. Smaller radii will result in sparser graphs, while larger radii will capture broader relationships. Similarly, the number of neighbors $k$ controls the sparsity (or density) of a $k$-NN graph. The relative number of edges for these two approaches will vary depending on the spatial distribution of the nodes. Additionally, we note that in a $k$-NN graph, a fixed number of nearest neighbors are chosen for each track, even if multiple neighbors have the same distance measure close to the track (many tracks can originate from the same point when their separation is lower than the resolution of the tracker). In contrast, radius graphs consider all these tracks regardless of their multiplicity, which offers a more accurate representation of the event structure.

\begin{figure}[t!]
	\centering
	\includegraphics[scale=0.55]{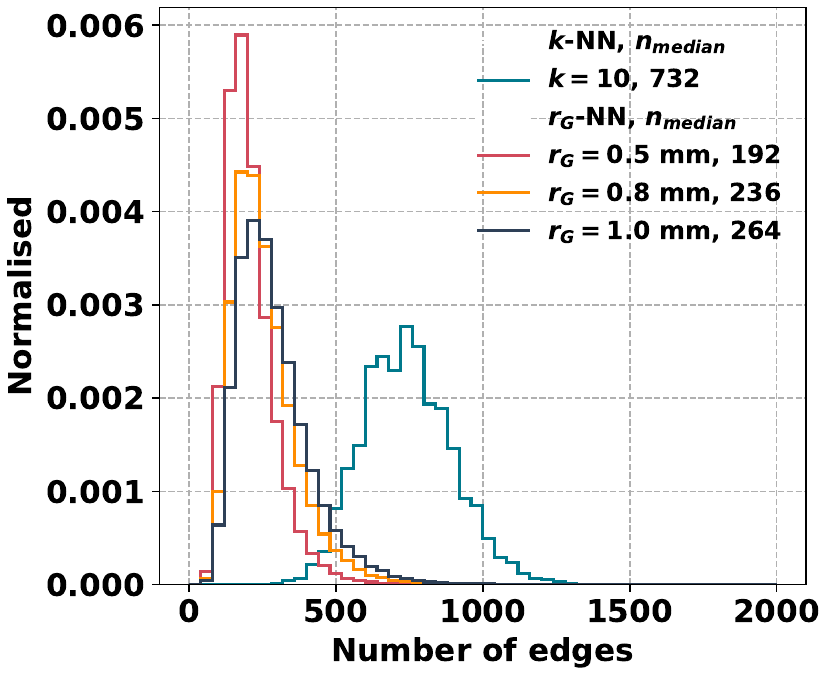} ~~
	\caption{ Number of edges in a graph for minbias events, constructed by considering neighbors within a specific radius ($r_G$) or by selecting a fixed number of neighbors using the $k$-nearest neighbors method. }
	\label{fig:nedegs}
\end{figure}

In analyzing events at L1, where computational efficiency is a crucial factor, the sparsity of the graph is an important factor in determining the graph construction algorithm. To this end, we look into the distribution of edge multiplicity for radius graphs and a relative $k$-NN construction. The histogram in Figure \ref{fig:nedegs} shows the difference in the number of edges between these two methods for various values of $r_{G}$ and \( k = 10 \) in minbias events.
 As we can see, the median number of edges for radius graphs and graphs constructed using the $k$-NN method differ considerably, with the former being significantly sparser. Therefore, we prefer to choose radius graphs over $k$-NN graphs in our study as they capture event structure better and have less computational overhead.

Each node is associated with ten node features. These features include the track's transverse momentum ($p_T$), momentum ($p$), transverse impact parameter ($d_0$), vertex position in cartesian coordinates (x, y, and z), outer position in the tracker in $\eta-\phi$ plane at the radial distance of 1.16 m, the distance of the closest approach to the beam line in the z-direction ($z_0$) and radial distance from the beamline ($D_T$). We additionally construct three edge attributes for every edge: the spatial separation along the x, y, and z axes between the connected nodes, the momentum and transverse momentum ratio between two linked nodes.

\begin{figure}[hbt!]
    \centering
    \includegraphics[scale=0.65]{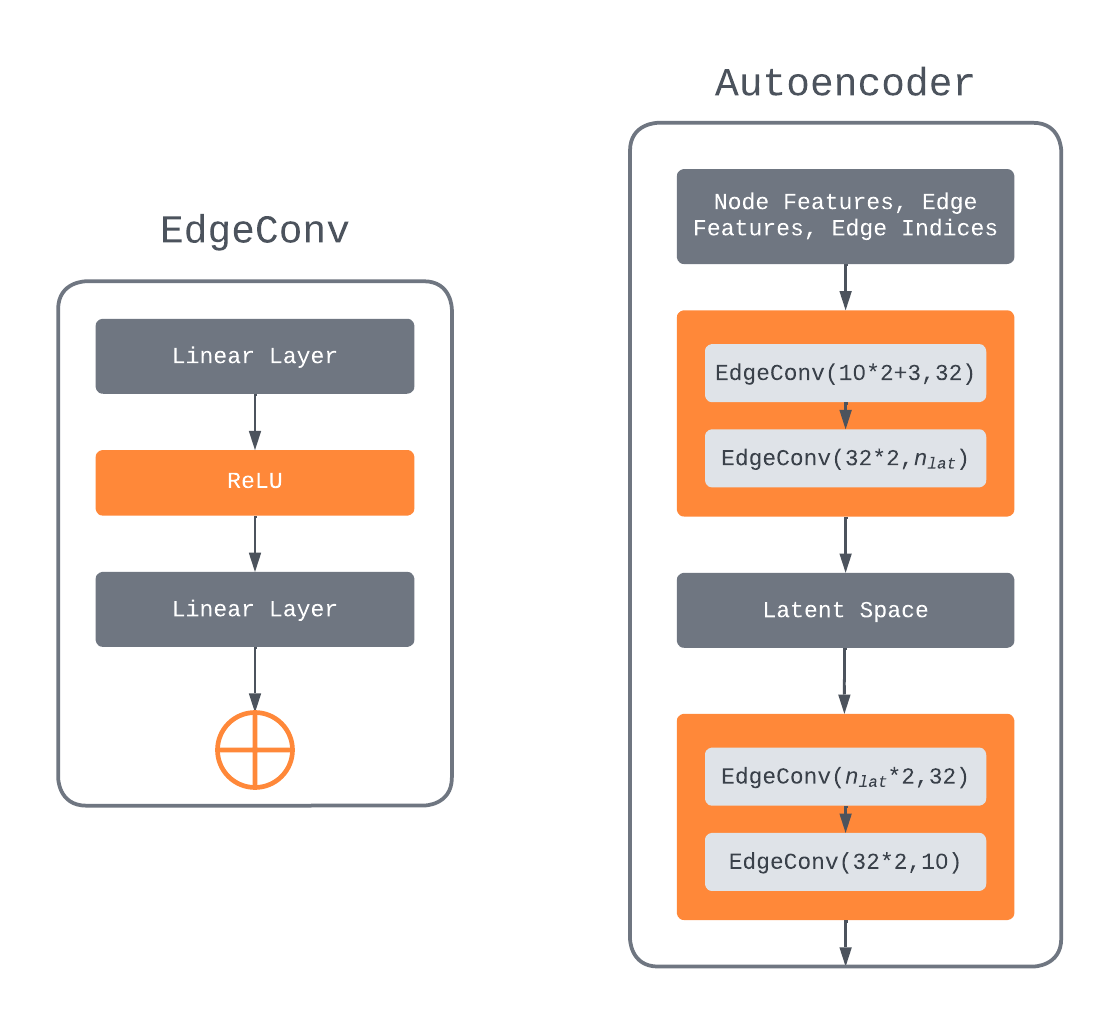} ~~

    \caption{The figure shows a schematic diagram of $\texttt{LLPNet}$, with latent node dimensions $n_{lat}$.}
    \label{fig:arch}
\end{figure}

The network architecture of the autoencoder employed in the present study is depicted in Figure \ref{fig:arch}. Autoencoders are generic neural network models where the input features undergo compression to a latent representation via an encoder. This latent representation again gets mapped to the same dimensions as the input features via a decoder, and the whole encoder-decoder setup is trained to reconstruct the input features faithfully via a loss function like mean squared error. In anomaly detection methods, the autoencoder is trained to reconstruct the most probable samples in the data (background processes at the LHC). In contrast, any sample (possible signals) with inherently different features from the training data would lead to a poor reconstruction, leading to larger values of the loss function.

In the autoencoder architecture used in this work, both encoder and decoder have two \texttt{EdgeConv} blocks where each \texttt{EdgeConv} block takes the node features and edge features (if present) and undergoes a message-passing operation, as illustrated in Figure \ref{fig:arch} (left) with both linear layers having an equal number of perceptrons to the specified updated node feature dimension. With the input node features and three edge features, the first \texttt{EdgeConv} operation takes a 23-dimensional input and produces 32-dimensional updated node features. The second \texttt{EdgeConv} layer accepts a 64-dimensional input (32 * 2) and leads to a $n_{lat}$-dimensional latent space where \(n_{lat}\) is a hyperparameter which we optimize as explained in the following sections. On the decoder side, the initial \texttt{EdgeConv} layer receives a 2*$n_{lat}$-dimensional input from the latent space and maps it to a 32-dimensional output. The second decoder layer, taking a 64-dimensional input (32 * 2), returns the reconstructed node features for each event. We use ``max'' aggregation operation ($\Delta$) for updating the nodes in all message-passing operations. With its relatively fewer learnable parameters ($\approx 6000$ at latent space dimension, $n_{lat} = 6$), this network architecture demonstrates suitability for deployment within the L1 stage of the trigger system

We train the autoencoder using QCD di-jet events with parton level $p_T$ ranging from 30 to 50 GeV, which includes 140 PU events. Rather than solely relying on minbias events for training, we incorporate the QCD di-jet events as the hard interaction with the PU for training the network. This approach enables us to capture simultaneously the characteristics of the hard QCD interaction and minbias events present in each event in terms of PU. On the other hand, an admixture of minbias samples with QCD di-jet events would lead to inefficient learning of the hard background features since the relative proportion (based on the cross-section) of QCD di-jet events would be much lower. Moreover, the harder $p_T$ bins of QCD di-jet events have relatively much lower cross-sections than the first bin but have similar features. For training the autoencoder, we utilize 600k events for training and 300k for validation, with 50k events taken for inference. For optimization, we employ the \texttt{ADAM} optimizer with a learning rate of 0.001 aided by a reduce-on-plateau condition with a 0.5 decay factor and patience of three epochs. To calculate the anomaly score for each event, we use the mean squared error (MSE) between the reconstructed event and the original event.

\section{Results}
\label{sec:results}

\begin{figure}[hbt!]
    \centering
    \includegraphics[scale=0.85]{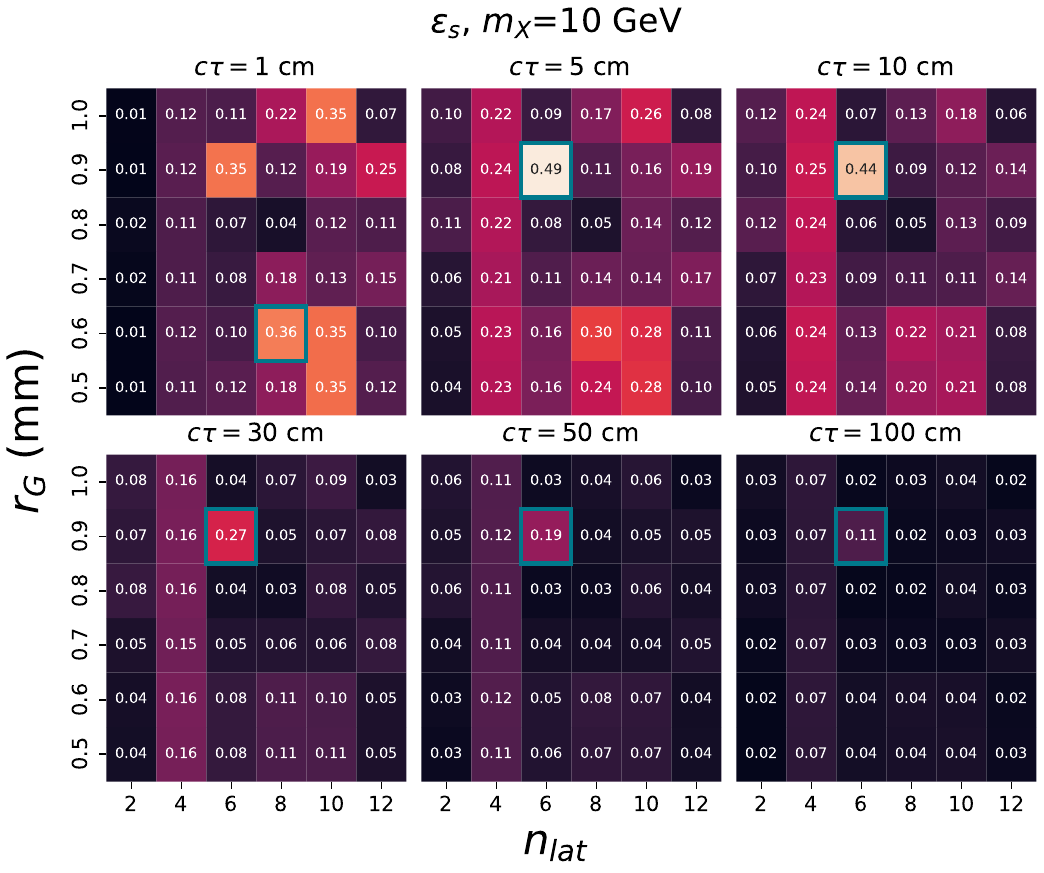} ~~
    \caption{Variation of signal efficiency at 99.9\% minbias background rejection ($\epsilon_{s}$) with graph radius ($r_{G}$) and latent space dimension ($n_{lat}$) for various $c\tau$ values of LLP with mass, $m_X=$ 10 GeV.}
    \label{fig:radiuscan_10}
\end{figure}

\begin{figure}[hbt!]
    \centering
    \includegraphics[scale=0.85]{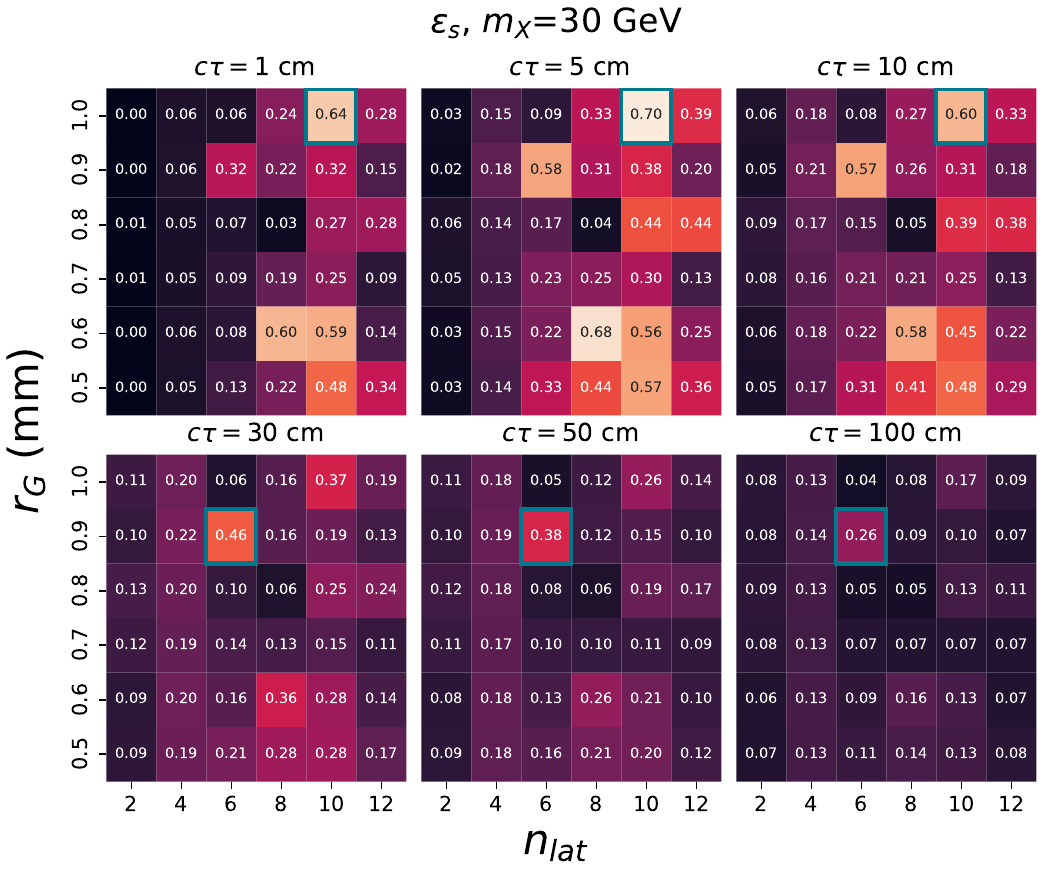} ~~
    \caption{Variation of signal efficiency at 99.9\% minbias background rejection ($\epsilon_{s}$) with graph radius ($r_{G}$) and latent space dimension ($n_{lat}$) for various $c\tau$ values of LLP with mass, $m_X=$ 30 GeV.}
    \label{fig:radiuscan_30}
\end{figure}

\begin{figure}[hbt!]
    \centering
    \includegraphics[scale=0.85]{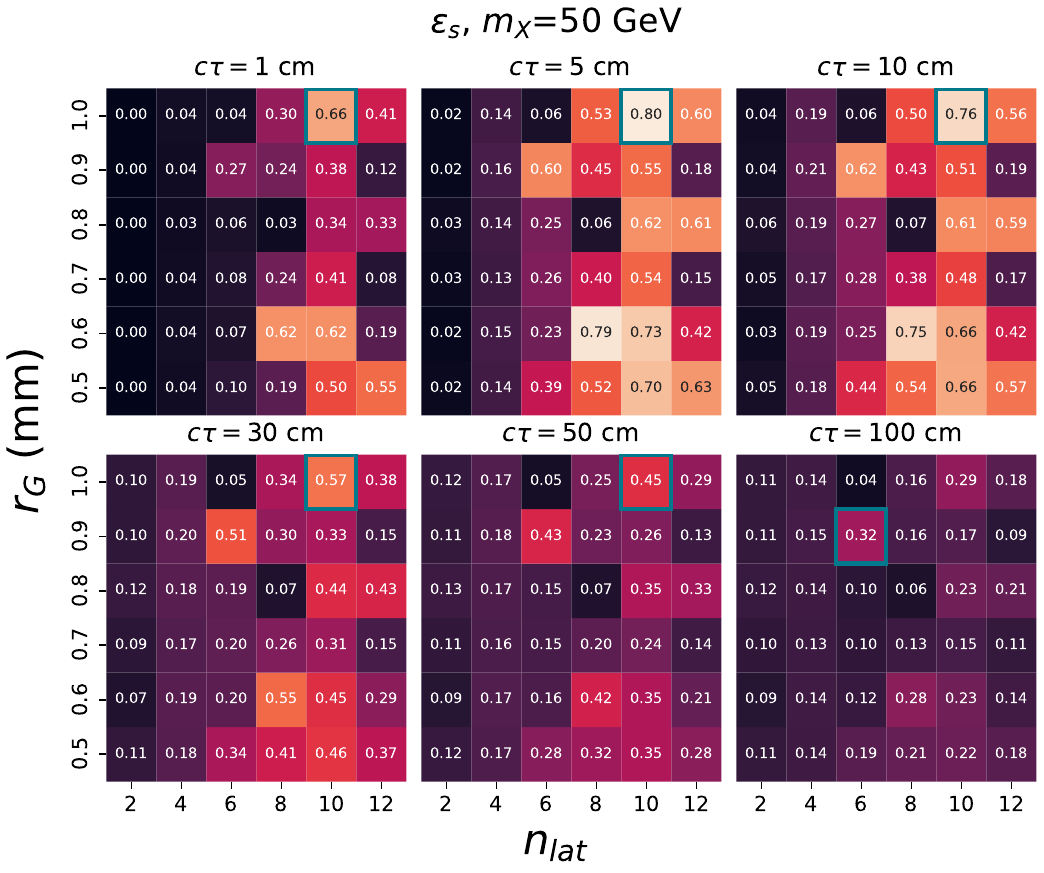} ~~
    \caption{Variation of signal efficiency at 99.9\% minbias background rejection ($\epsilon_{s}$) with graph radius ($r_{G}$) and latent space dimension ($n_{lat}$) for various $c\tau$ values of LLP with mass, $m_X=$50 GeV.}
    \label{fig:radiuscan_50}
\end{figure}

We evaluate the autoencoder using two key metrics: the Area Under the Curve (AUC) for the Receiver Operator Characteristics (ROC) and the signal efficiency when the background rejection rate is 99.9\% (denoted as $\epsilon_s$). The choice of a 99.9\% background rejection rate in the second metric takes into consideration the order of magnitude of background rejection needed at L1, which is on the order of $10^3$. The values for AUC and $\epsilon_s$ are calculated from a dataset with eighteen benchmark points for LLPs and the minbias background. Each of these benchmarks, including the minbias sample, contains 140 PU events. The LLP benchmarks are characterized by masses, $m_X=$ 10, 30, and 50 GeV and $c\tau$ that span from 1 cm to 100 cm. Our analysis includes a comprehensive scan of hyperparameters, specifically the graph radius, $r_G$, and dimension of the latent space, $n_{lat}$. Figures \ref{fig:radiuscan_10}, \ref{fig:radiuscan_30}, and \ref{fig:radiuscan_50} illustrate how $\epsilon_s$ changes with $c\tau$ across different $r_G$ and $n_{lat}$ for LLP masses of 10, 30, and 50 GeV respectively. We vary the $r_G$ from 0.5 to 1 mm and the $n_{lat}$ from 2 to 12. We also highlight the configurations that maximize $\epsilon_s$. Corresponding plots with AUC values are provided in the Appendix \ref{app:roc_varyingR}.

As seen in Figures \ref{fig:radiuscan_10}, \ref{fig:radiuscan_30}, and \ref{fig:radiuscan_50}, there's no single common point in latent space and graph radius that yields the maximum signal efficiency $\epsilon_{s}$ for every LLP benchmark point. Therefore, we must identify optimal points that provide the best possible signal efficiency for LLPs. This includes both very low mass LLPs ($m_X=$ 10 GeV) and relatively heavy mass LLPs ($m_X=$ 50 GeV), as well as LLPs with short decay lengths ($c\tau=$ 1 cm) and those with large decay lengths ($c\tau=$ 100 cm).

To identify the optimal combination of $n_{lat}$ and $r_{G}$ that maximize the $\epsilon_{s}$ for our anomaly detector, we employed the Gradient Boosting Regressor (GBR). Our dataset was structured such that combinations of $n_{lat}$ and $r_G$ served as features, and their corresponding $\epsilon_{s}$ acted as the target. The GBR, with parameters set at \texttt{`n\_estimators=1500'}, \texttt{`learning\_rate=0.01'}, \texttt{`max\_depth=6'}, and \texttt{`loss=squared\_error'}, was trained on this dataset. Post training, the model predicted the $\epsilon_{s}$  for all potential combinations, and the one yielding the highest predicted $\epsilon_{s}$ was considered the optimal hyperparameter set. The optimization was performed separately for different masses. Subsequently, we adopted a similar approach for different decay lengths.

\begin{figure}[hbt!]
    \centering
    \includegraphics[scale=0.365]{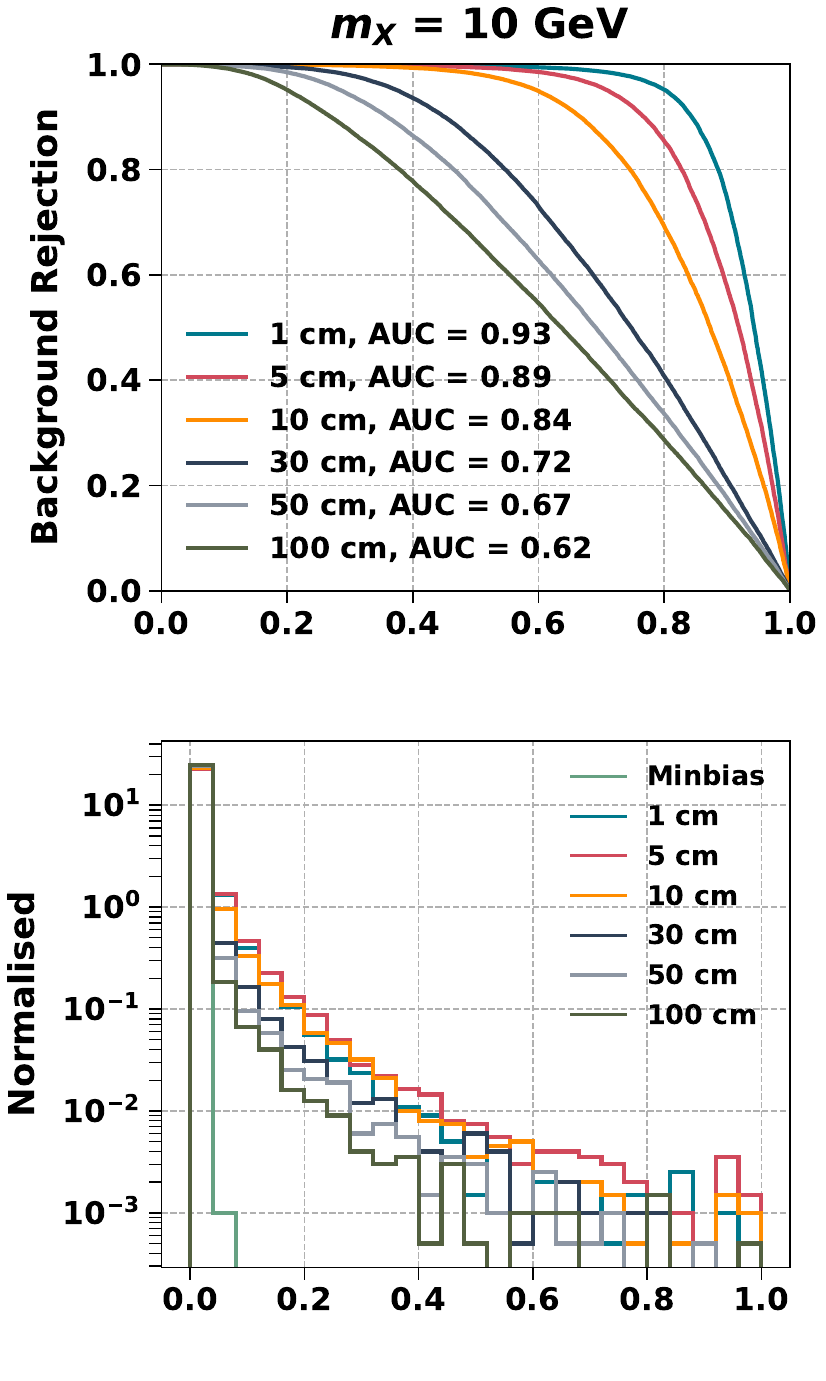}%
    \includegraphics[scale=0.365]{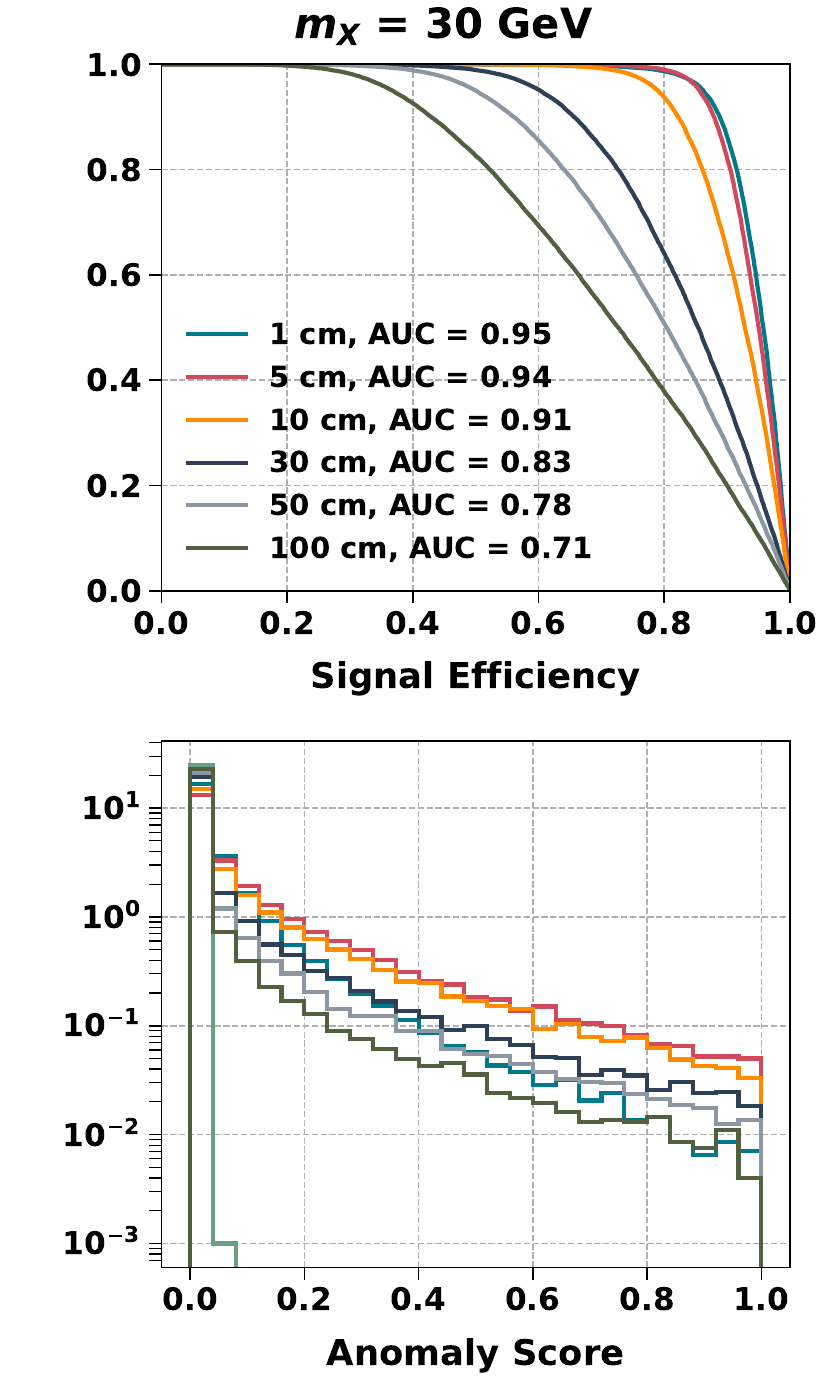}%
    \includegraphics[scale=0.365]{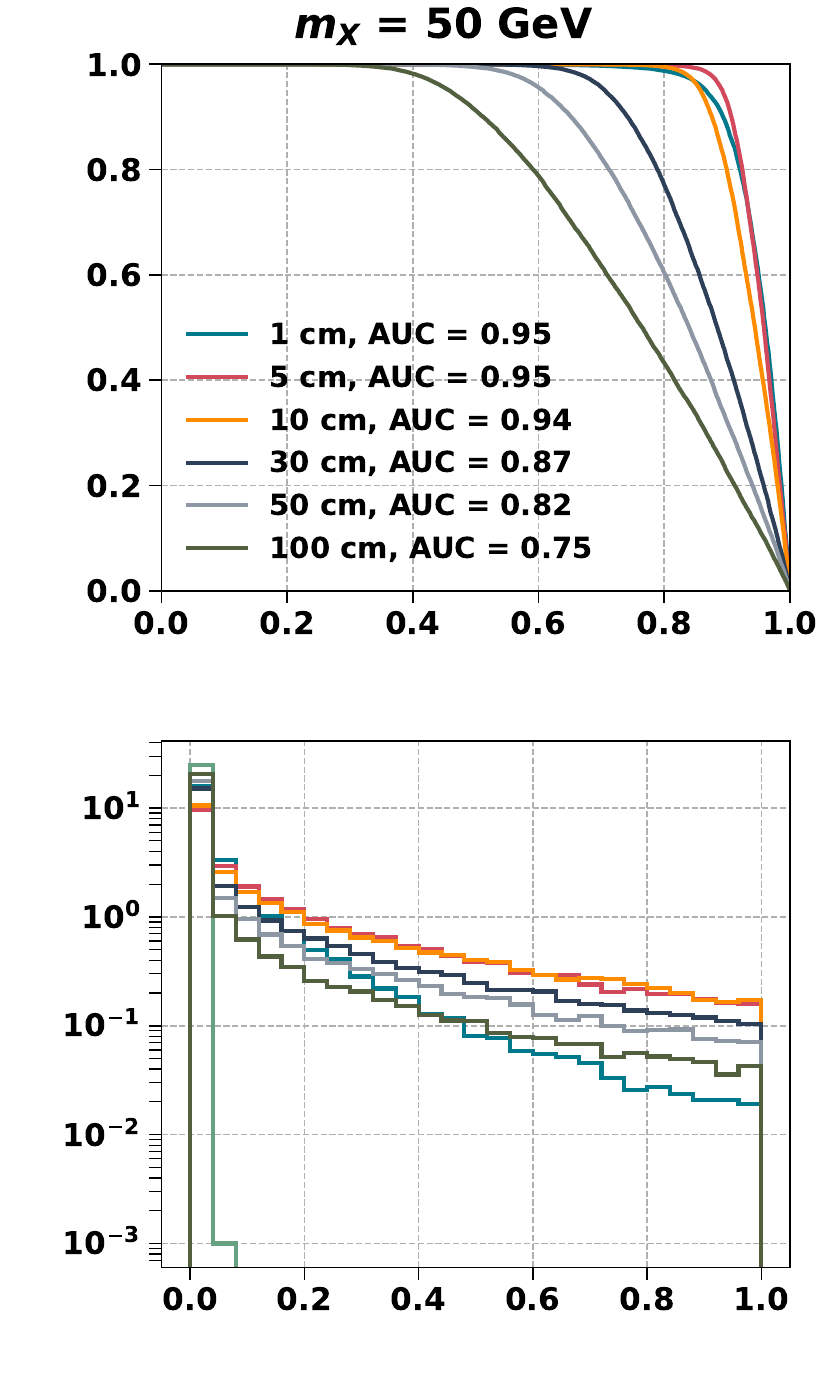} 

    \caption{ ROC curves (\textit{top}), along with anomaly scores (\textit{bottom}), for minbias and specific LLP benchmark points with $m_X=$ 10 GeV (\textit{left}), 30 GeV (\textit{middle}), and 50 GeV (\textit{right}) for WP-1. The decay length of LLP ranges from 1 cm to 100 cm.}
    \label{fig:roc_anomaly_min_wp1}
\end{figure}

 Table \ref{tab:masses} presents the optimal $r_G$ and $n_{lat}$ for three distinct LLP masses, $m_X=$10, 30, and 50 GeV. 
\begin{table}[h!]
\centering
\begin{tabular}{ccc}
\toprule
$m_{X}$ (GeV) & $r_{G}$ (mm) & $n_{lat}$ \\
\midrule
10 & 0.9 & 6 \\
30 & 1 & 10 \\
50 & 1 & 10 \\
\bottomrule
\end{tabular}
\caption{Optimal parameters (graph radius, $r_G$,  and latent space dimension, $n_{lat}$) for different LLP masses ($m_{X}$).}
\label{tab:masses}
\end{table}
Similarly, Table \ref{tab:decay_lengths} enumerates the same parameters but for various decay lengths ranging from 1 cm to 100 cm. 
\begin{table}[h!]
\centering
\begin{tabular}{ccc}
\toprule
$c\tau$ (cm) & $r_G$ (mm) & $n_{lat}$ \\
\midrule
1 & 1 & 10 \\
5 & 0.6 & 8 \\
10 & 0.9 & 6 \\
30 & 0.9 & 6 \\
50 & 0.9 & 6 \\
100 & 0.9 & 6 \\
\bottomrule
\end{tabular}
\caption{Optimal parameters (graph radius, $r_G$,  and latent space dimension, $n_{lat}$) for different LLP decay lengths ($c\tau$).}
\label{tab:decay_lengths}
\end{table}
These tables present the results of a hyperparameter scan, highlighting the configurations that yielded the best performance for each LLP scenario. From these tables, it becomes clear that the optimal performance for each LLP benchmark can be reached using one of the following three combinations in graph radius and latent space dimensions:  [0.6, 8], [0.9, 6] and [1, 10]. Thus, we have identified three working points (WPs) corresponding to these values, referred to as WP-1, WP-2, and WP-3. In Figure \ref{fig:roc_anomaly_min_wp1} (\textit{top}), we explicitly show the ROC curve for WP-1 with $r_G = 0.6$ mm and $n_{lat}=$ 8 for above mentioned LLP scenarios. In Figure \ref{fig:roc_anomaly_min_wp1} (\textit{bottom}), the anomaly scores for each of these LLP benchmark points, along with the minbias background, are presented explicitly. Similar plots for WP-2 and WP-3 are included in Appendix \ref{app:roc_wp2wp3}.

\subsection{Background rate calculation}
\label{sec:rate}
The QCD samples across different $p_T^{gen}$ bins, as explained in Section \ref{sec:sigback}, are combined with minbias events to calculate the event rate using ``stitching" method as explained in \cite{Ehataht:2021rkh} and validated in one of our previous studies \cite{Bhattacherjee:2021qaa}. When calculating the background rate for triggers by combining different $p_T^{gen}$ bins can lead to unavoidable overlap in phase space. The stitching method helps to overcome this issue by applying selected weights to the simulated events. It also lets us gradually increase the number of simulated events in specific areas of phase space ($p_T^{gen}$) if the existing MC samples don't cover those areas well.
The calculation of the event weight in terms of rate is performed using the following formula-
\begin{equation}
w^{I} = \frac{F}{N_{incl}+\sum_j N_j \times \frac{n_j}{(N_{PU}+1)\times p_j}}
\label{eq:stitch}
\end{equation}
Here, $F$ is the $pp$ collision frequency (around 28 MHz\footnote{Although proton beams cross each other every 25 ns yet only 70\% of these intersections result in \(pp\) collisions.}). $N_{incl}$ refers to the total number of events with $N_{PU}$+1 minbias events, where $N_{PU}$ is the average PU events in a collision. $N_j$ is the event count for the $j^{th}$ $p_T^{gen}$ bin, and $n_j$ is the number of inelastic $pp$ interactions in that bin. $p_j$ is the probability of a single inelastic collision in the $j^{th}$ $p_T^{gen}$ bin, determined by comparing the cross-section for that bin to the cross-section without any $p_T^{gen}$ conditions.

The technical design report for the L1 trigger at the CMS for the HL-LHC outlines the design of two dedicated triggers tailored towards triggering jets originating from LLPs with short and large decay lengths. The combined allowable rate bandwidth for these triggers is approximately 40 KHz so we need to keep the background rate below 40 KHz. For the current study, we constrain the background rate at 30 KHz. 

\begin{figure}[hbt!]
    \centering
    \includegraphics[scale=0.44]{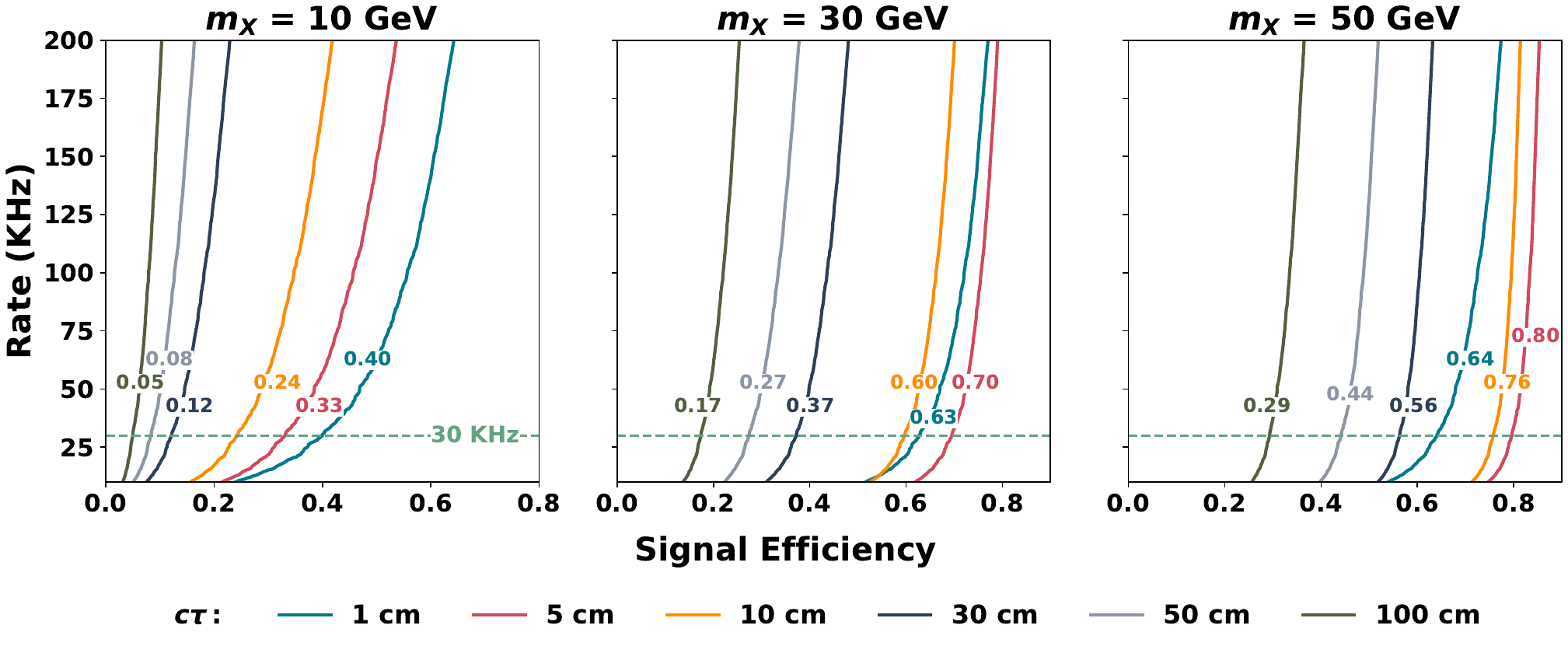} 
    \caption{ Background rate versus signal efficiency for LLPs with mass, $m_X=$ 10 GeV, 30 GeV and 50 GeV for decay lengths, $c\tau$, varying from 1 to 100 cm for WP-1. Annotated values correspond to signal efficiency at background rate of 30 KHz.}
    \label{fig:rate_wp1}
\end{figure}

\begin{figure}[hbt!]
    \centering
    \includegraphics[scale=0.44]{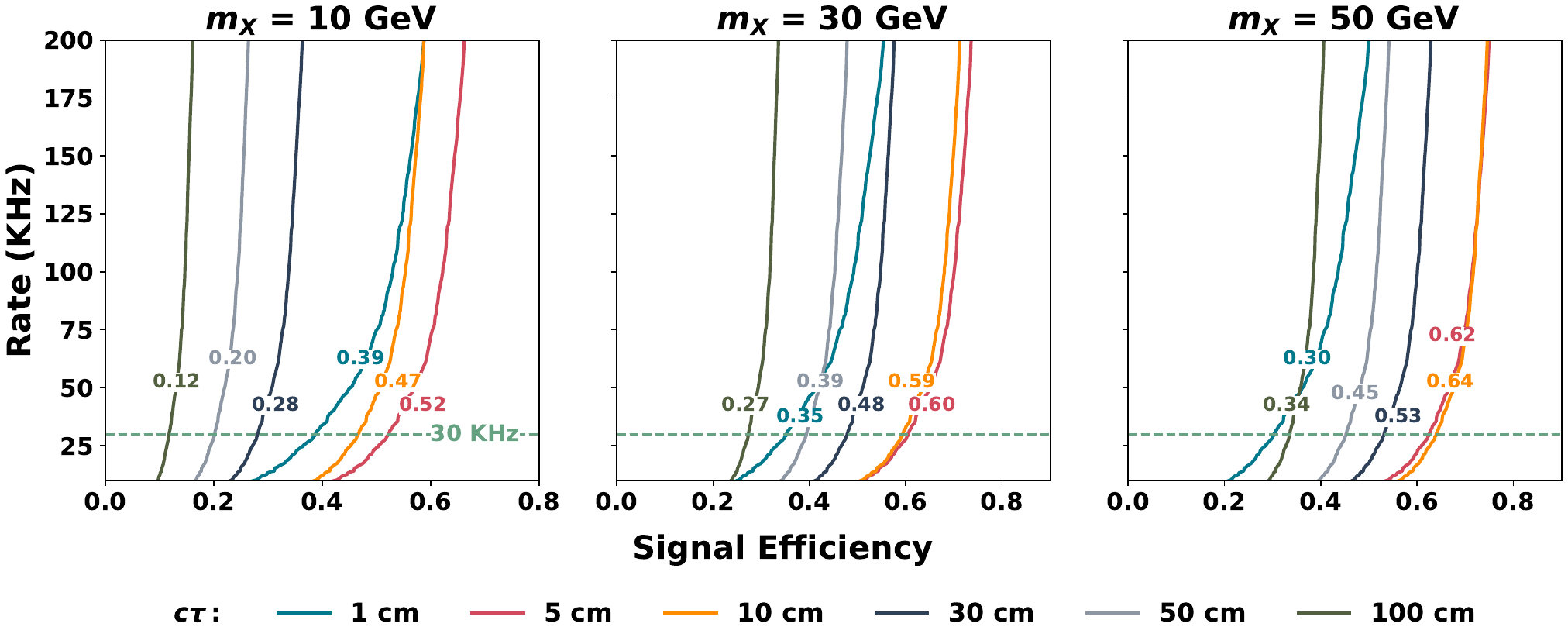} 
    \caption{ Background rate versus signal efficiency for LLPs with mass, $m_X=$ 10 GeV, 30 GeV and 50 GeV for decay lengths, $c\tau$, varying from 1 to 100 cm for WP-2. Annotated values correspond to signal efficiency at background rate of 30 KHz.}
    \label{fig:rate_wp2}
\end{figure}

\begin{figure}[hbt!]
    \centering
    \includegraphics[scale=0.44]{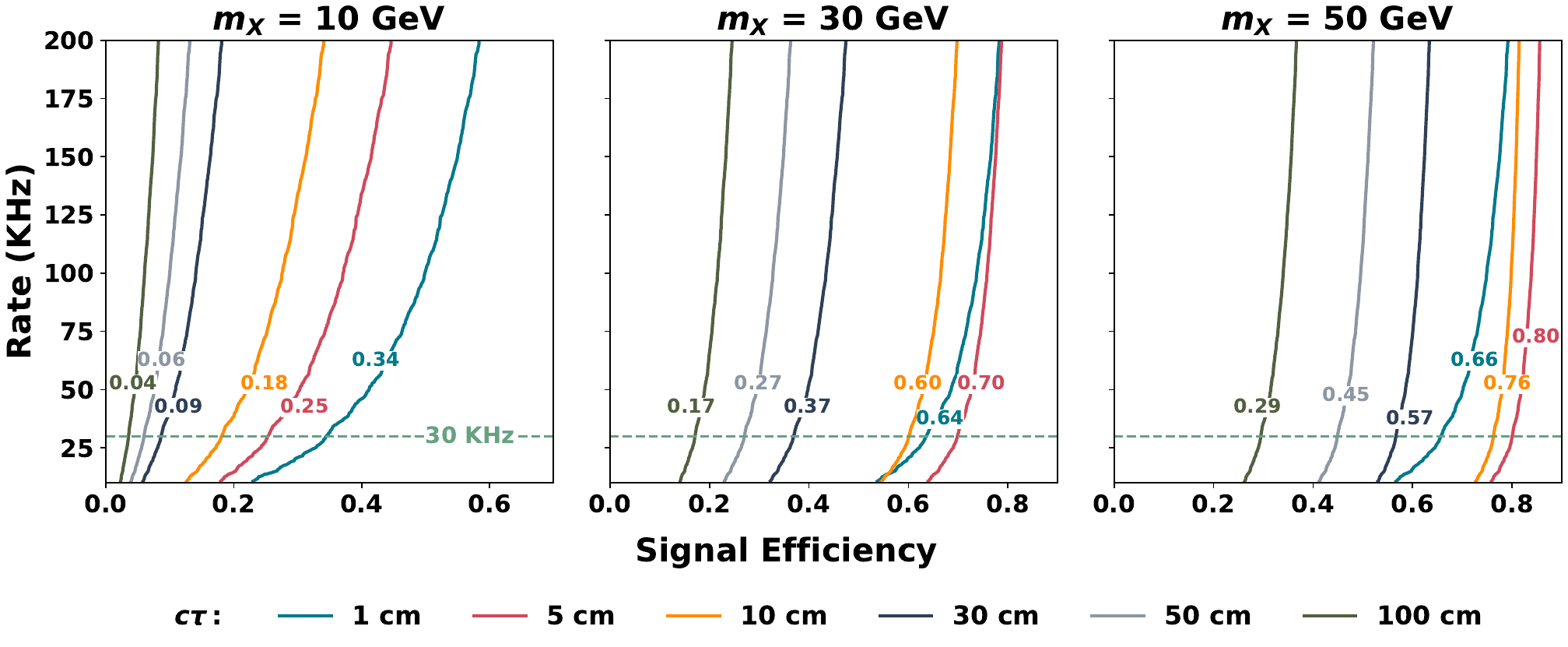} 
    \caption{ Background rate versus signal efficiency for LLPs with mass, $m_X=$ 10 GeV, 30 GeV and 50 GeV for decay lengths, $c\tau$, varying from 1 to 100 cm for WP-3. Annotated values correspond to signal efficiency at background rate of 30 KHz.}
    \label{fig:rate_wp3}
\end{figure}

In Figure \ref{fig:rate_wp1}, \ref{fig:rate_wp2}, and \ref{fig:rate_wp3} the variation of trigger rate is plotted against signal efficiency for three LLP benchmark points for WP-1, WP-2 and WP-3 respectively, with LLP masses ($m_{X}$) of 10, 30, and 50 GeV. Within these points, the decay length is considered across a range from 1 cm to 100 cm.

 In these plots, the signal efficiency for various benchmark points is annotated at a background rate of 30 KHz. In Table \ref{tab:sige_wp}, we also enumerate the signal efficiency for various signal benchmark points for LLPs with $m_X=$ 10, 30 and 50 GeV and $c\tau$ varying from 1 cm to 100 cm for three WPs: WP-1, WP-2 and WP-3 after constraining the background rate at 30 KHz. 
 
 \begin{table}[hbt!]
    \centering
    \begin{tabular}{cS[table-format=0.2]S[table-format=0.2]S[table-format=0.2]S[table-format=0.2]S[table-format=0.2]S[table-format=0.2]}
        \toprule
        & \multicolumn{6}{c}{$c\tau$ (cm)} \\
        \cmidrule(lr){2-7}
        \multirow{-2}{*}{$m_X$ (GeV)} & {1} & {5} & {10} & {30} & {50} & {100} \\
        \midrule
        \multicolumn{7}{c}{\cellcolor{lightgray}WP-1} \\
        10 & 0.40 & 0.33 & 0.24 & 0.12 & 0.08 & 0.05 \\
        30 & 0.63 & 0.70 & 0.60 & 0.37 & 0.27 & 0.17 \\
        50 & 0.64 & 0.80 & 0.76 & 0.56 & 0.44 & 0.29 \\
        \midrule
        \multicolumn{7}{c}{\cellcolor{lightgray}WP-2} \\
        10 & 0.39 & 0.52 & 0.47 & 0.28 & 0.20 & 0.12 \\
        30 & 0.35 & 0.60 & 0.59 & 0.48 & 0.39 & 0.27 \\
        50 & 0.30 & 0.62 & 0.64 & 0.53 & 0.45 & 0.34 \\
        \midrule
        \multicolumn{7}{c}{\cellcolor{lightgray}WP-3} \\
        10 & 0.34 & 0.25 & 0.18 & 0.09 & 0.06 & 0.04 \\
        30 & 0.64 & 0.70 & 0.60 & 0.37 & 0.27 & 0.17 \\
        50 & 0.66 & 0.80 & 0.76 & 0.57 & 0.45 & 0.29 \\
        \bottomrule
    \end{tabular}
    \caption{Signal efficiency values for different LLP signal benchmark points across three working points (WP-1, WP-2, WP-3) at background rate of 30 KHz.}
    \label{tab:sige_wp}
\end{table}
 Table \ref{tab:sige_wp} illustrates that good signal efficiency is achievable for LLPs with both higher and lower decay lengths. As a result, a single anomaly-based trigger might be sufficient to trigger the LLPs, regardless of whether the decay length is short or large. For example, for LLPs with a decay length of 5 cm, signal efficiencies of approximately 52\%, 60\%, and 62\% can be achieved for LLPs with $m_X=$ 10 GeV, 30 GeV, and 50 GeV with WP-2, respectively. At a decay length of 100 cm, the signal efficiency retains respectable values at around 12\%, 27\%, and 34\% for LLPs with $m_X=$ 10 GeV, 30 GeV, and 50 GeV respectively. 
 
For context, when considering a decay length of 5 cm for $m_X$= 30 GeV resulting from the decay of a 125 GeV Higgs boson, the CMS experiment estimates a signal efficiency of roughly 8\%. This is evaluated at a background rate of 30 KHz in a 200 PU scenario, utilizing the displaced tracker $H_T$ trigger. This trigger is designed for LLPs with shorter decay lengths at HL-LHC \cite{CERN-LHCC-2020-004}. Meanwhile, for a $c\tau$ value of 100 cm and $m_X$= 50 GeV, the CMS experiment projects a signal efficiency close to 20\% at a background rate of 25 KHz in a 200 PU environment using the displaced Calo-Jet trigger, primarily designed for LLPs with a significant lifetime at HL-LHC \cite{CERN-LHCC-2020-004}.

\section{Summary and conclusion}
\label{sec:summary}

Triggering exotic events within the level-1 (L1) trigger system continues to be a substantial challenge for many searches involving displaced particles. High pile-up (PU) environment at HL-LHC will further add to this challenge making it exceedingly difficult to trigger such events efficiently at L1 while keeping the background rate under control. In scenarios where long-lived particles (LLPs) are coupled to the Standard Model (SM)-like 125 GeV Higgs boson, the triggering of such events is primarily dependent on the associated production modes of the Higgs boson where these searches predominantly depend on triggering events with objects produced in association with the Higgs boson. However, this approach comes with a reduced production cross-section, making the search for such particles exceedingly challenging.
In the specific scenario where the Higgs boson is SM-like with a mass of 125 GeV, a further challenge arises. The hadronic activity resulting from the decay of LLPs, which originate from the decay of the Higgs boson, is very small. This low level of hadronic activity further complicates the triggering process, making detection more difficult using standard $H_T$ or jet triggers. Therefore, there is an immediate and pressing requirement to develop L1 triggers capable of independently capturing LLP events, without relying on existing triggers. 

With the introduction of advanced FPGAs and forthcoming upgrades to the triggering and DAQ system at HL-LHC, the implementation of machine learning, particularly graph-based machine learning triggers, becomes viable due to relaxed latency constraints at L1 and reliable and efficient deployment of such networks on FPGAs.  In this work, we study a novel method to trigger displaced events using a machine learning approach where we utilize message passing autoencoder employing ``Edge convolution'' on the tracks reconstructed at L1 where these collection of tracks act as point cloud.  

In consideration of the low latency requirement at L1, the network architecture of the autoencoder is deliberately designed to be simple, featuring a comparatively small number of learnable parameters. For the LLP benchmark points, we focus on the LLPs with mass of 10, 30, and 50 GeV and decay lengths extending from 1 cm to 100 cm. Each LLP is assumed to decay into two displaced jets, with a branching ratio of 100\%, resulting in a final state marked by displaced jet signatures. The kinematic features of these tracks, along with their displacement in the transverse plane, are utilized to construct the node and edge features for event-based graphs. Instead of the $k$-nearest neighbours method, we adopt the radius method to form these graphs. This approach better captures the event structure and offers a reduced computational load, a crucial aspect for L1 triggers. A hyperparameter scan is conducted over the graph radius and latent space dimensions. Subsequently, based on regression analysis with the ``Gradient Boosting Regressor" performed for each mass and decay length independently, three distinct working points (WPs) are established. However, it is of significance to acknowledge that the optimal radius of the graphs employed in the experiments may vary depending upon the attainable spatial track resolution at HL-LHC, taking into account factors such as the track \(p_T\) and position within the tracker along with the mean number of PU events.  

Results are presented in terms of background rate versus signal efficiency. The background rate computation employs the ``Stitching method", taking into consideration both minbias and QCD di-jet backgrounds taking into account the potential overlaps in phase space. QCD background is segmented into multiple $p_T^{gen}$ bins to ensure precise rate calculations. We observe notable signal efficiency for signal benchmark points at the permissible background rate of 30 KHz. Although, signal efficiency decreases with increase in decay length as expected, it retains a respectable value for higher decay lengths. For example, we observe signal efficiency of 33\%, 70\% and 80\% for LLPs of mass 10, 30 and 50 GeV with WP-1. At decay length of 50 cm, signal efficiency of 20\%, 39\% and 45\% is observed for 10, 30 and 50 GeV LLP with WP-2. 

In conclusion, the graph-based autoencoder highlighted in this study suggests a potential for enhancing signal acceptance for light LLPs at the L1 trigger level for HL-LHC. However, it is imperative to underscore that additional comprehensive studies are warranted to evaluate the overall performance and suitability of such autoencoders, especially in relation to FPGA implementations while carefully considering the inherent L1 latency constraints.

\section*{Acknowledgement}
B.B. acknowledges the support provided by the MATRICS Grant (MTR/2022/000264) from the Science and Engineering Research Board (SERB), Government of India. P.S. thanks Rhitaja Sengupta for valuable discussions. Additionally, P.S. expresses gratitude to the Indian Institute of Science (IISc), Bengaluru, and the Physical Research Laboratory (PRL), Ahmedabad, for facilitating P.S.'s visit to PRL for scientific collaboration. This work utilized the Param Vikram-1000 High-Performance Computing Cluster and TDP resources at the PRL and computing facility at the Centre for High Energy Physics (CHEP), IISc.
\clearpage
\appendix

\section*{Appendix}
\section{AUC values in radius vs. latent plane}
\label{app:roc_varyingR}
\begin{figure}[hbt!]
    \centering
    \includegraphics[scale=0.85]{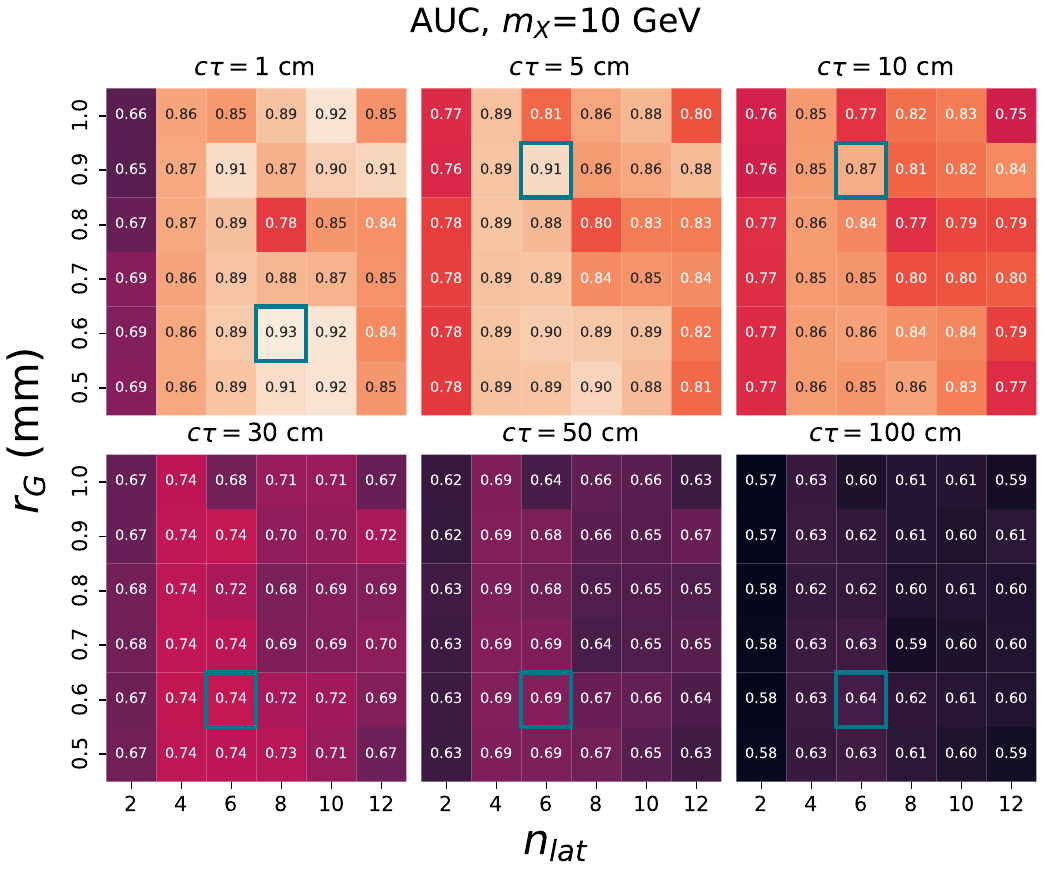} ~~
    \caption{Variation of area under curve (AUC) with graph radius ($r_{G}$) and latent space dimension ($n_{lat}$) for various $c\tau$ values of LLP with mass, $m_X=$ 10 GeV.}
    \label{fig:scan_10}
\end{figure}

\begin{figure}[hbt!]
    \centering
    \includegraphics[scale=0.85]{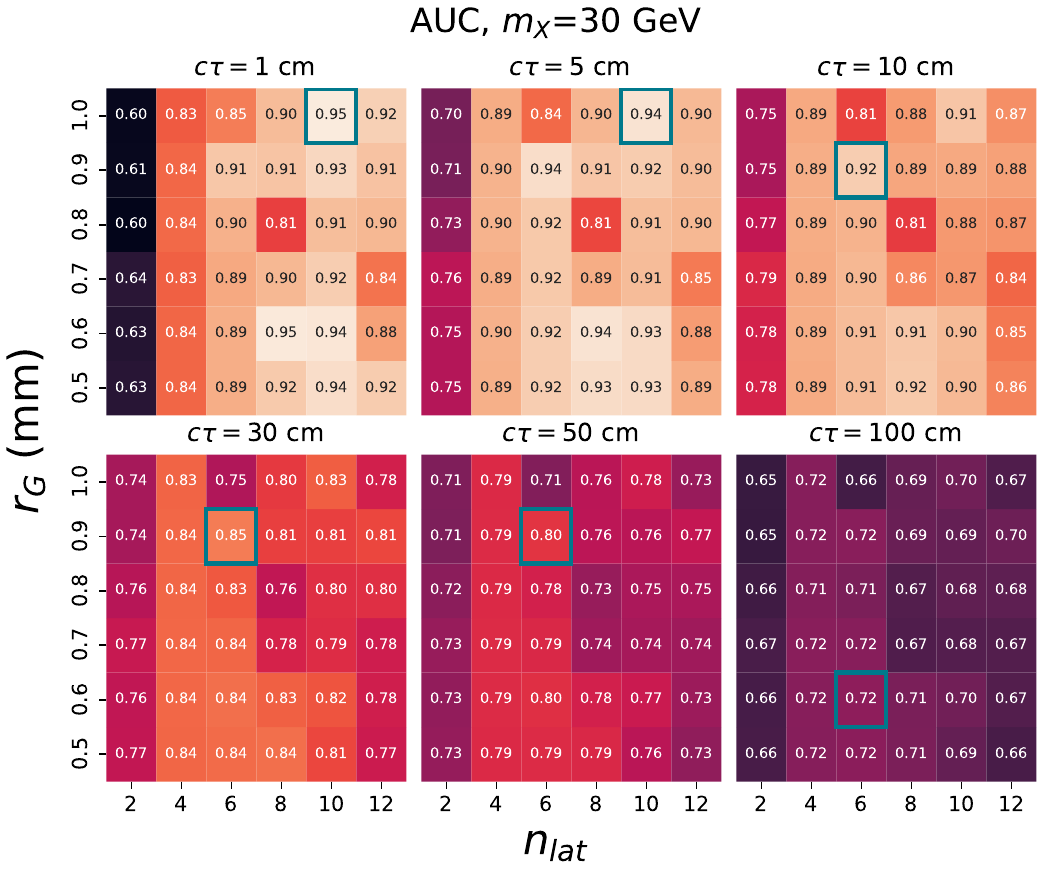} ~~
    \caption{Variation of area under curve (AUC) with graph radius ($r_{G}$) and latent space dimension ($n_{lat}$) for various $c\tau$ values of LLP with mass, $m_X=$ 30 GeV.}
    \label{fig:scan_30}
\end{figure}

\begin{figure}[hbt!]
    \centering
    \includegraphics[scale=0.85]{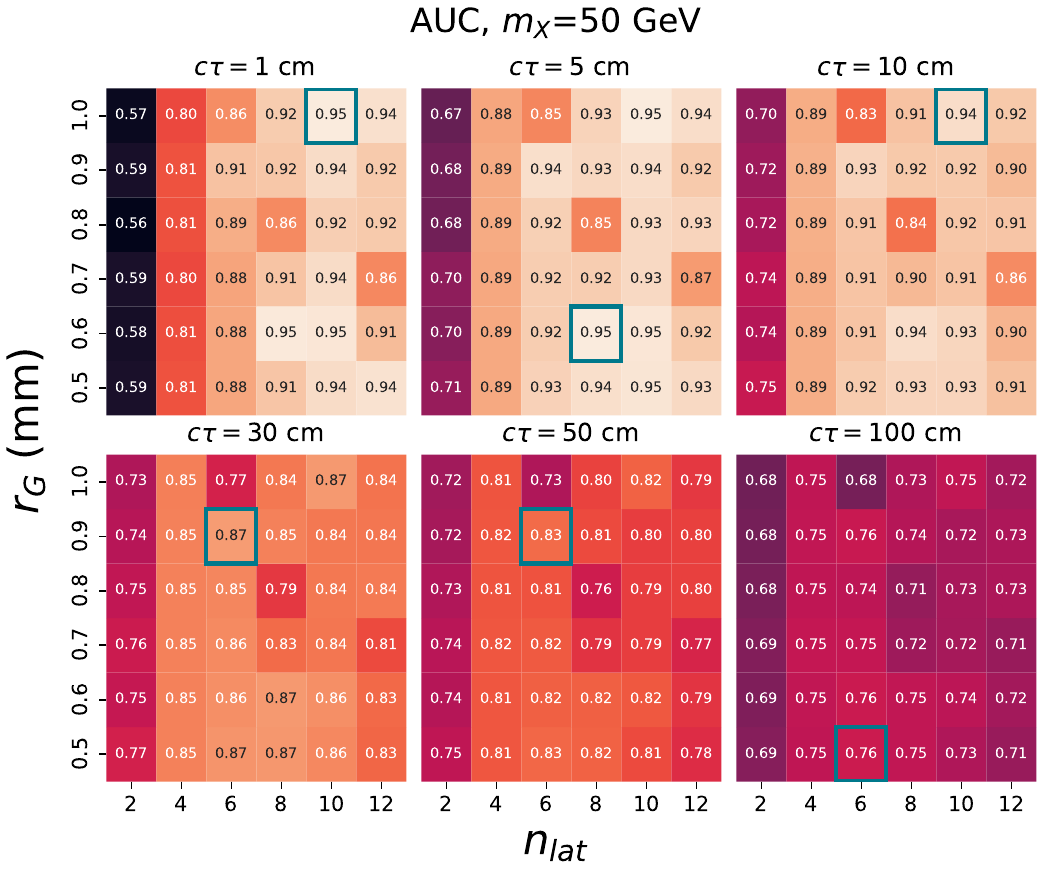} ~~
    \caption{Variation of area under curve (AUC) with graph radius ($r_{G}$) and latent space dimension ($n_{lat}$) for various $c\tau$ values of LLP with mass, $m_X=$ 50 GeV.}
    \label{fig:scan_50}
\end{figure}

Figure \ref{fig:scan_10}, \ref{fig:scan_30} and \ref{fig:scan_50} represents the area under curve (AUC) values of ROC for combinations of graph radius ($r_G$) and latent space dimension ($n_{lat}$) for LLPs of mass, $m_X=$ 10, 30 and 50 GeV for decay length, $c\tau$, varying between 1 cm and 100 cm. Combination of $r_G$ and $n_{lat}$ with maximum AUC is highlighted. 

\clearpage

\section{ROC curves for WP-2 and WP-3}
\label{app:roc_wp2wp3}
\begin{figure}[hbt!]
    \centering
    \includegraphics[scale=0.35]{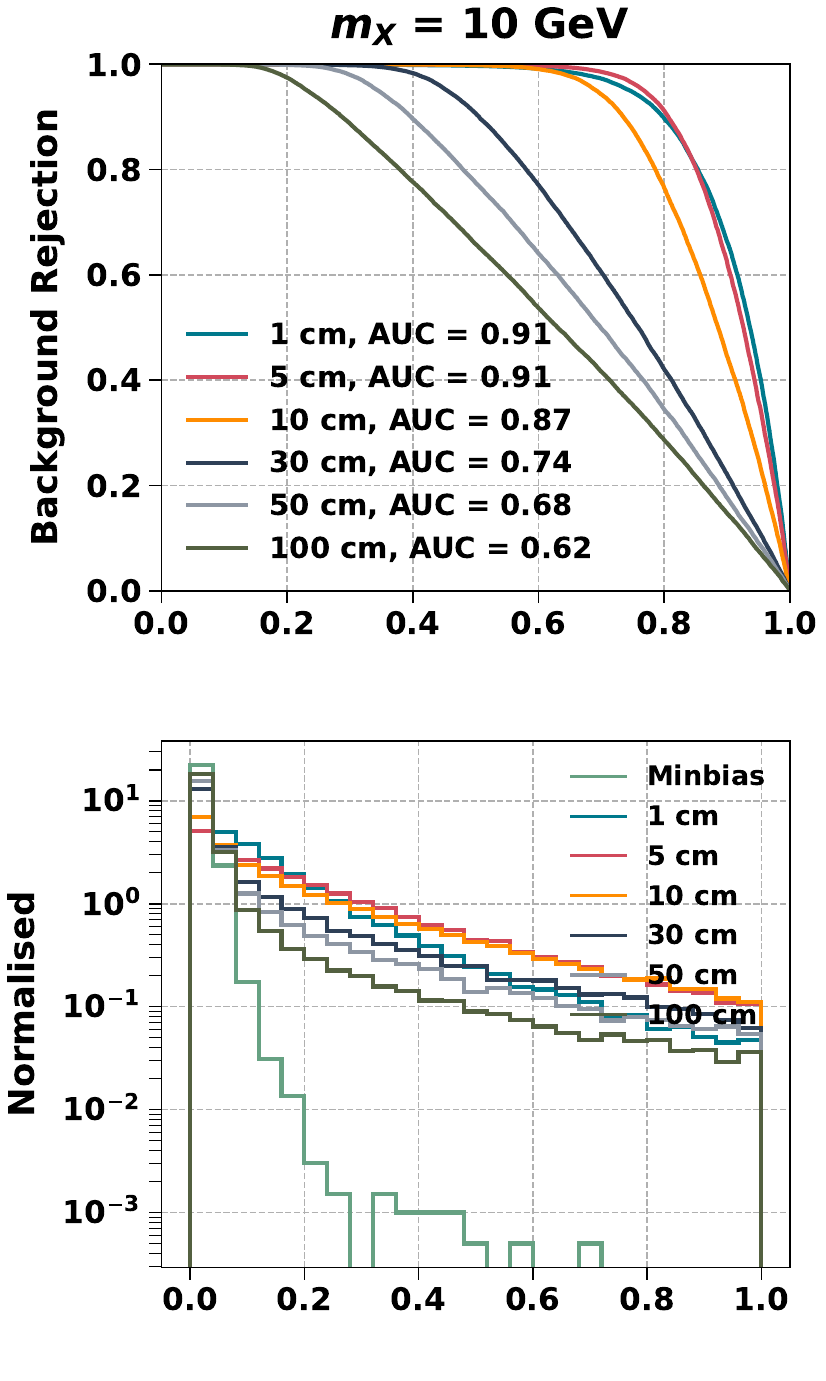}
    \includegraphics[scale=0.35]{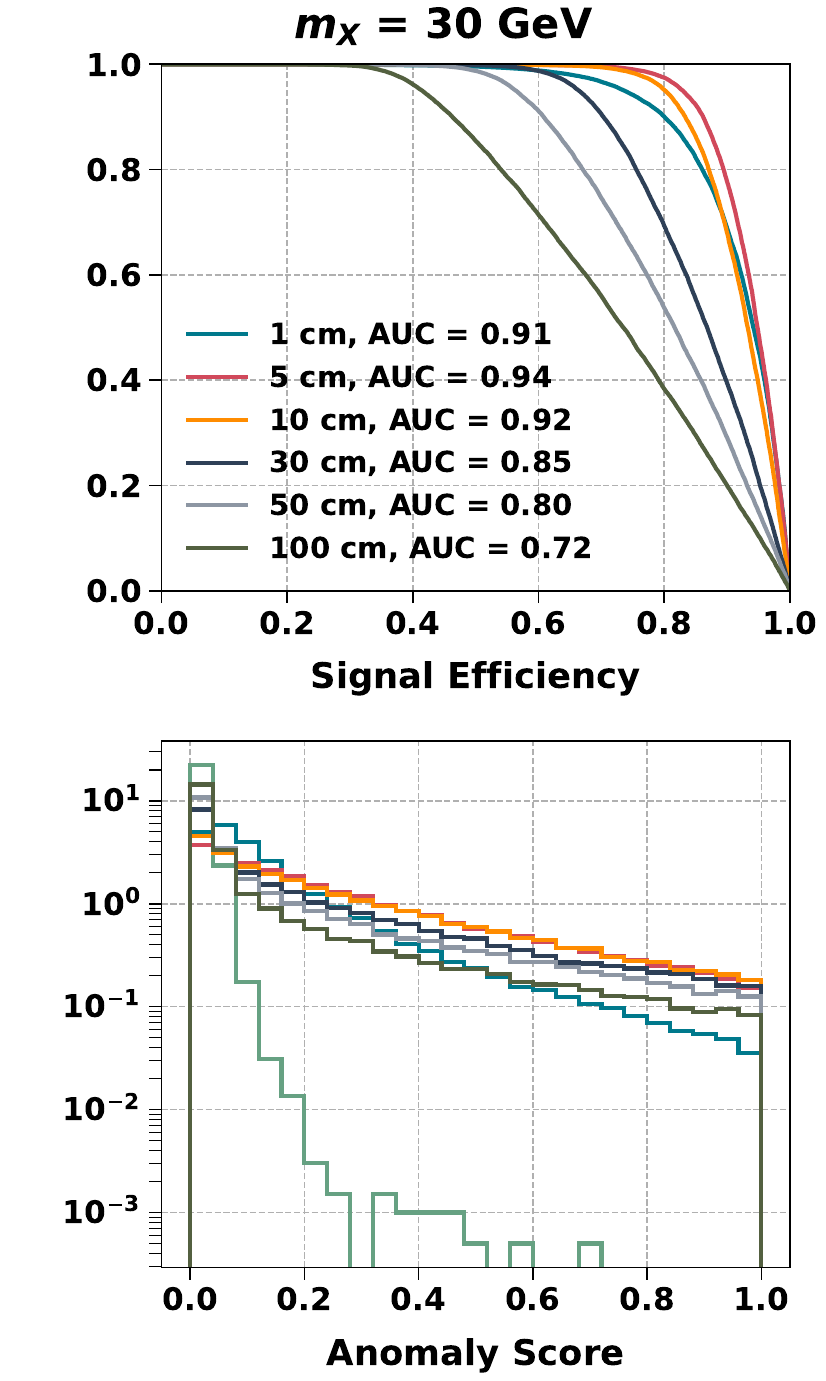}
    \includegraphics[scale=0.35]{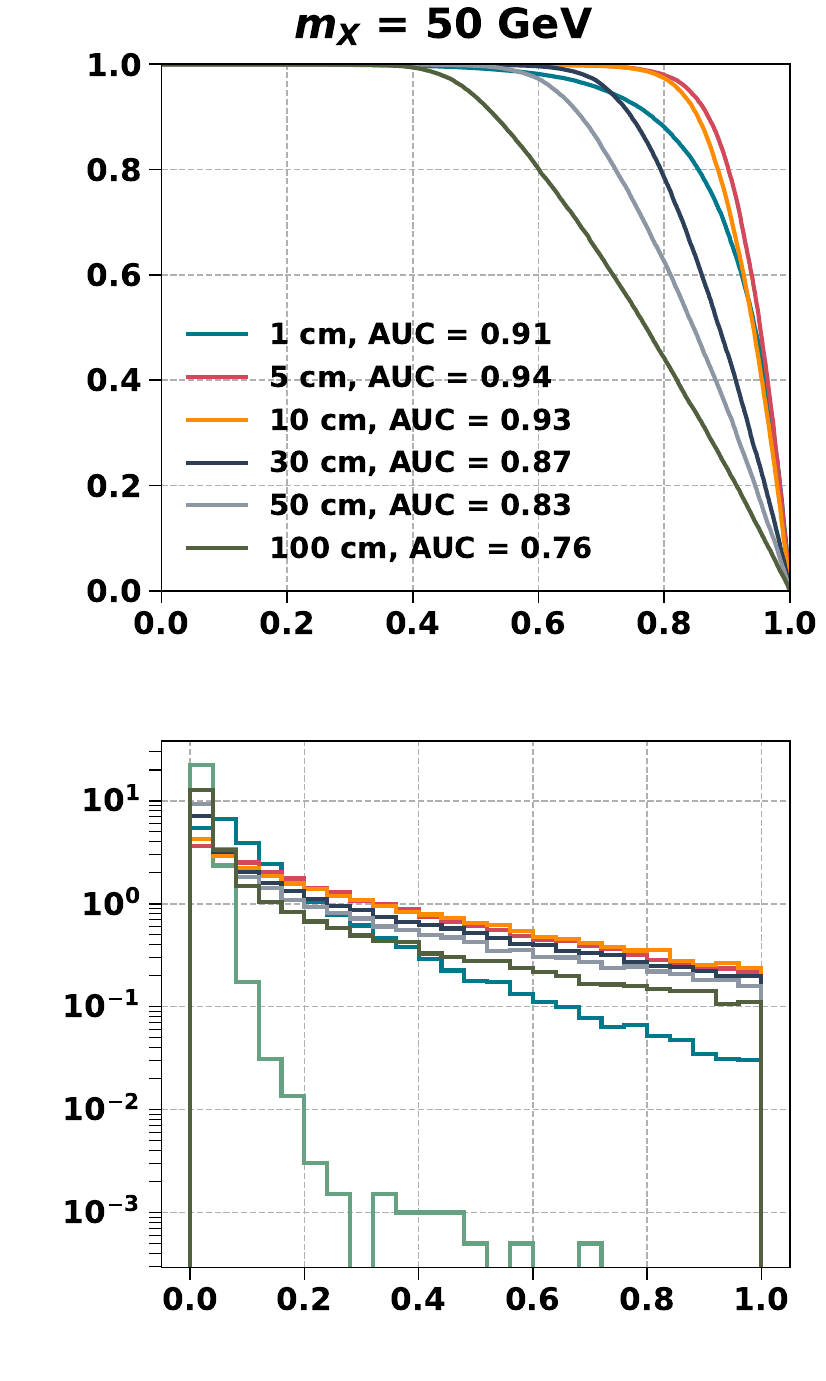}

    \caption{ ROC curves (\textit{top}), along with anomaly scores (\textit{bottom}), for minbias and specific LLP benchmark points with $m_X=$ 10 GeV (\textit{left}), 30 GeV (\textit{middle}), and 50 GeV (\textit{right}) for WP-2. The decay length of LLP ranges from 1 cm to 100 cm.}
    \label{fig:roc_anomaly_min_wp2}
\end{figure}

\begin{figure}[hbt!]
    \centering
    \includegraphics[scale=0.35]{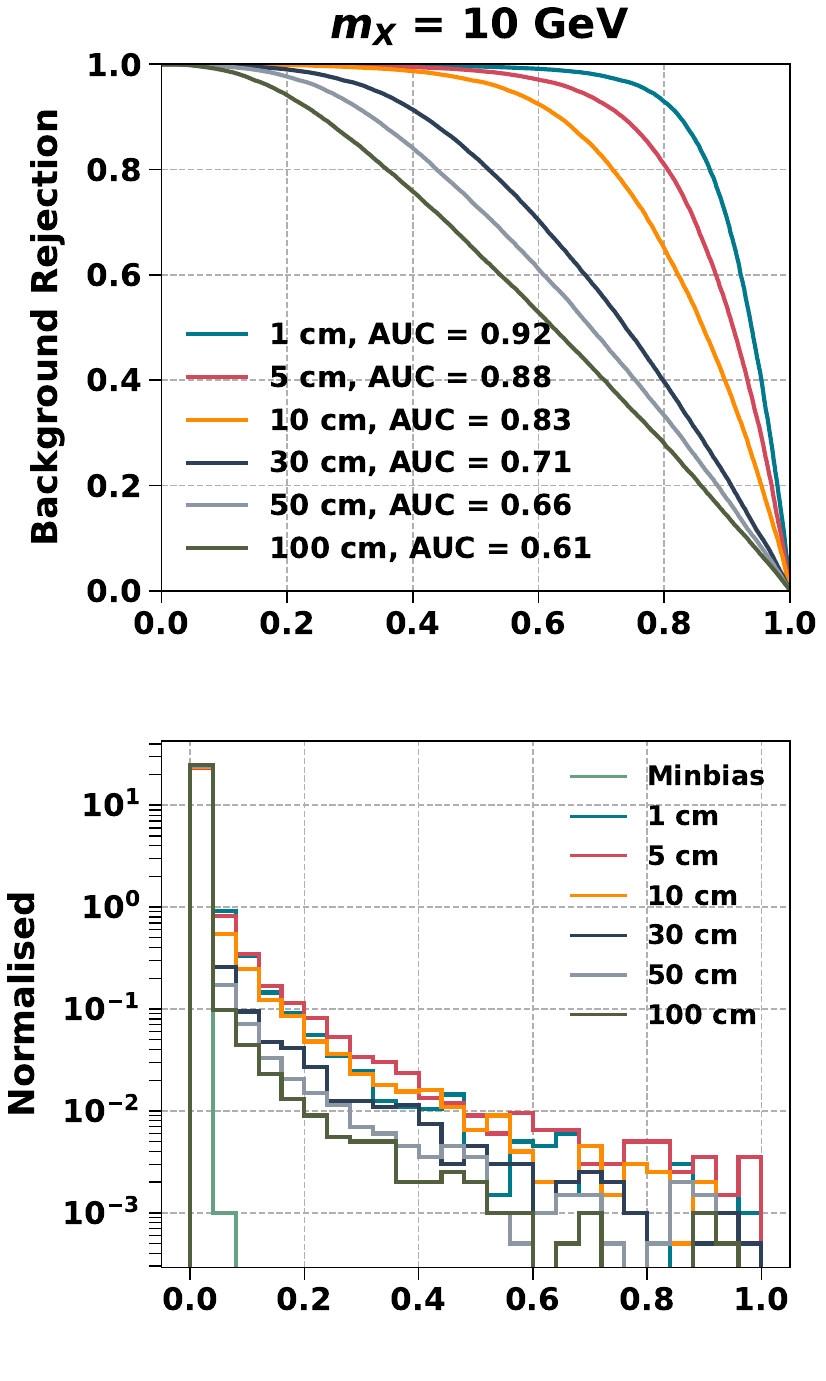}
    \includegraphics[scale=0.35]{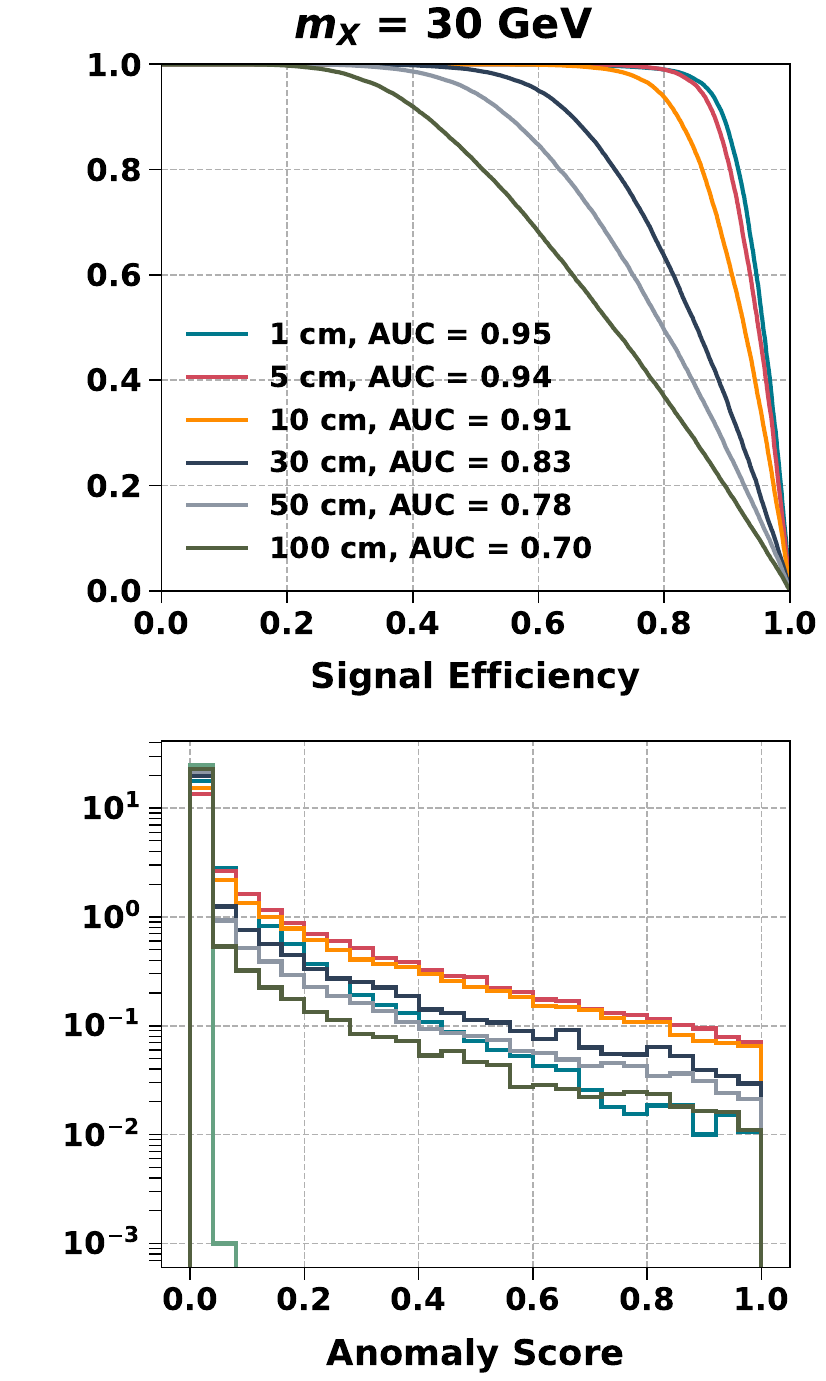}
    \includegraphics[scale=0.35]{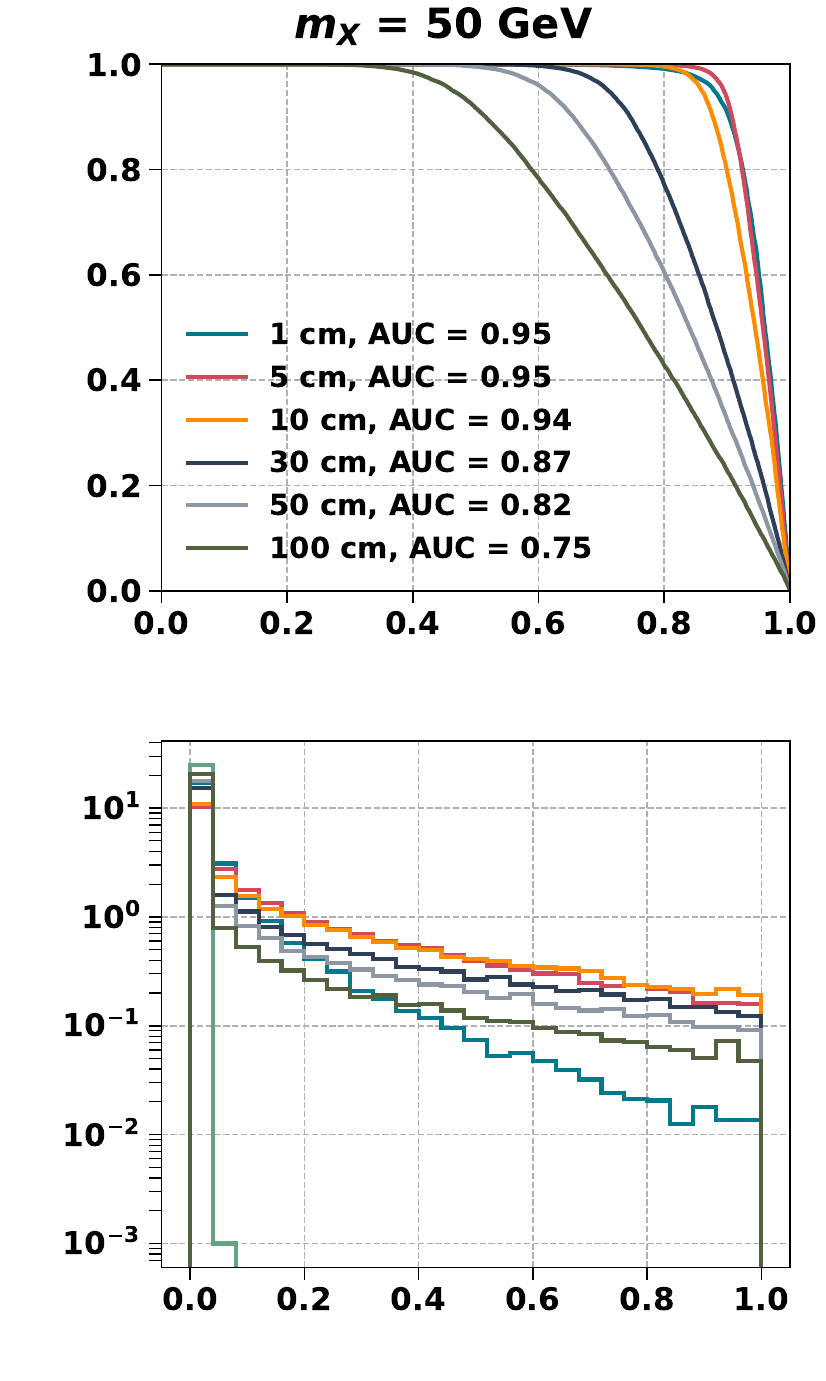}

    \caption{ ROC curves (\textit{top}), along with anomaly scores (\textit{bottom}), for minbias and specific LLP benchmark points with $m_X=$ 10 GeV (\textit{left}), 30 GeV (\textit{middle}), and 50 GeV (\textit{right}) for WP-3. The decay length of LLP ranges from 1 cm to 100 cm.}
    \label{fig:roc_anomaly_min_wp3}
\end{figure}

Figures \ref{fig:roc_anomaly_min_wp2} and \ref{fig:roc_anomaly_min_wp3} (\textit{top}) depict the ROC curve for WP-2 with $r_G = 0.9$ and $n_{lat}=$ 6 and WP-3 with $r_G = 1$ and $n_{lat}=$ 10 respectively for various LLP scenarios. In Figures \ref{fig:roc_anomaly_min_wp2} and \ref{fig:roc_anomaly_min_wp3} (\textit{bottom}), the anomaly scores for each of these LLP benchmark points, along with the minbias background, are presented explicitly.

\clearpage
\newpage
\bibliographystyle{JHEP.bst}
\providecommand{\href}[2]{#2}\begingroup\raggedright\endgroup
\end{document}